# Decorative Plasmonic Surfaces

Hamid T. Chorsi, Ying Zhu, and John X. J. Zhang

***Abstract*— Low-profile patterned plasmonic surfaces are synergized with a broad class of silicon microstructures to greatly enhance near-field nanoscale imaging, sensing, and energy harvesting coupled with far-field free-space detection. This concept has a clear impact on several key areas of interest for the MEMS community, including but not limited to ultra-compact microsystems for sensitive detection of small number of target molecules, and "surface" devices for optical data storage, micro-imaging and displaying. In this paper, we review the current state-of-the-art in plasmonic theory as well as derive design guidance for plasmonic integration with microsystems, fabrication techniques, and selected applications in biosensing, including refractive-index based label-free biosensing, plasmonic integrated lab-on-chip systems, plasmonic near-field scanning optical microscopy and plasmonics on-chip systems for cellular imaging. This paradigm enables low-profile conformal surfaces on microdevices, rather than bulk material or coatings, which provide clear advantages for physical, chemical and biological-related sensing, imaging, and light harvesting, in addition to easier realization, enhanced flexibility, and tunability.**

***Index Terms*—Microsystems, microelectromechanical Systems, plasmonic surface, micro/nano fabrication, plasmonics, biosensing.**

## I. Introduction

Recent advances in micro-nanoscale technology and in the understanding of the mechanisms of light interaction with miniature structures have opened a wide range of possibilities in manipulating, tailoring, transmitting, and processing near-field optical signals in the same way that electric signals are successfully processed at lower frequencies [1-3]. Optical metamaterials [4-6] in particular exploit the resonant wave interaction of collections of plasmonic nanoparticles and are able to produce anomalous light interaction beyond naturally available optical materials. However, the implementation of these concepts in real microsystems are limited by a variety of factors, including technological challenges in realizing three-dimensional specific nano-structured patterns; inherent device sensitivity to fabrication induced disorder and losses; and experiment-guided modeling of structure interaction with photons across multiple scales.

Inspired by the concepts of optical metamaterials and the peculiar features of plasmonic nanopatterns, "Plasmonic microsystems" is an emerging field that is evolving into a novel paradigm for the conception of optical plasmonic surfaces. Plasmonic patterning is the 2-D sub-wavelength conformal arrangements of plasmonic nanoparticles, nanoantennas, and nanoapertures or nanogrooves to achieve unconventional optical wave interactions with nanoscale surface structures. This provides a large degree of flexibility and enhanced features in overall response in imaging and sensing.

It should be noted that standard solutions to pattern the optical radiation are already available in the form of lenses, reflectors and gratings [7-10]. However, these devices require relatively large sizes, do not offer easy tunability or flexibility to direct the beam, and do not provide emission or radiation enhancement because they are simply based on the reflection and refraction properties of standard materials. On the other hand, properly patterned plasmonic surfaces on current and future microelectromechanical (MEMS) devices may provide much enhanced *flexibility* in the available degrees of freedom to tailor the signal radiation, with additional features of interest, such as *tunability* of the angle of radiation, significant overall *emission enhancement,* and extremely *low profile*.

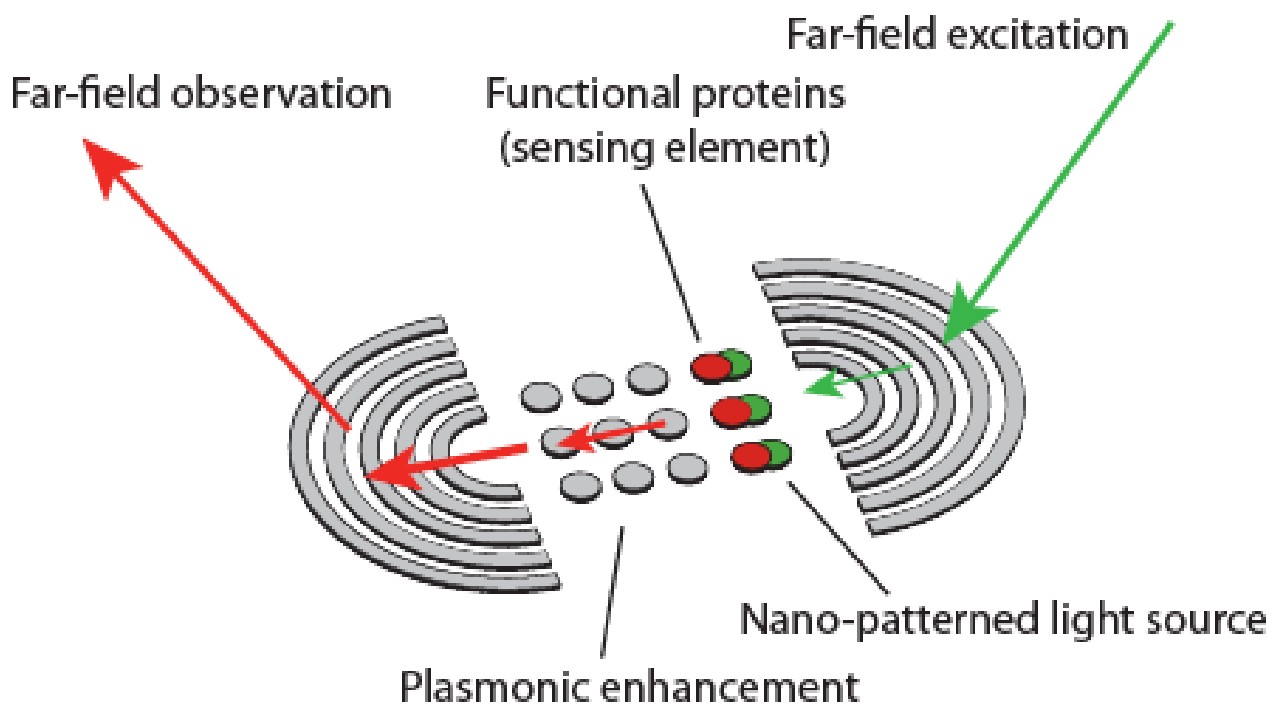

Fig. 1. Schematics of proposed concept: plasmonic microsystems surface geometries combining optical plasmonic patterns of different nature to realize the bridge between nanoscale sensing, imaging and far-field optical radiation.

This paragraph of the first footnote will contain the date on which you submitted your paper for review. It will also contain support information, including sponsor and financial support acknowledgment. For example, "This work was supported in part by the U.S. Department of Commerce under Grant BS123456".

H. T. Chorsi, Y. Zhu, and X.J. Zhang are with Thayer School of engineering, Dartmouth College, Hanover, NH 03755 USA. (e-mail: Hamid.chorsi@ucdenver.edu,Ying.Zhu@dartmouth.edu,John.Zhang@dartmouth.edu).

In plasmonic microsystems, we need to thoroughly explore novel designs and geometries to flexibly manipulate optical radiation and emission. This can be accomplished through optical metasurfaces composed of nanopatterned arrays or reflectors characterized by ultralow profile that are capable of enhancing light emission in the desired direction and tailoring the radiation pattern at will. These concepts may be applied to both transmission and reception, and have the potential to create a much needed efficient and flexible transition link between MEMS based nanoscale sensing and imaging in the near-field (Fig. 1) and radiation and detection in the far-field.

Implementing plasmonic microsystems involves the design, technology and fabrication efforts intended at merging electromagnetic waves with electromechanical functions. This technology provides a framework to combine current micro-scale devices with nano-scale plasmonic components exempt from the diffraction limit. In particular, plasmonic microsystems are formed when a nano-scale structure or particle interacts or combines with a micro-dimension device. Examples of plasmonic microsystems include gold nanoparticles (dimensions are usually less than 100 nm) on top of solid surface such as silicon wafer or glass slide, or nano-slit grooves on near-field scanning optical microscopy (NSOM) probes.

This paper is organized as follows: Section II reviews the fundamental theories and phenomena that establish the basis for the study of plasmonic microsystems. Section III describes recent advances in the fabrication of sub-100-nm patterns on micro scale devices and structures. Section IV reviews the recent developments in plasmonic microsystems for biosensing applications, including refractive-index based label-free biosensing, plasmonic integrated lab-on-chip systems, plasmonic for NSOM and plasmonics on-chip system for cellular imaging. Section V presents our perspective on the current direction in plasmonic micromachining. Section VI concludes this paper.

## II. Theory

Plasmonics is a field of science that explores plasmons; collective oscillations of free electron clouds excited via polarized waves. The field of plasmonics is key in exploring ultrafast and miniaturized electronic, mechanical, and biomedical devices. In this section, we first review the most important theories and phenomena that form the basis for the study of plasmonic waves. We then summarize the theoretical physics presented in recent works on plasmonic microsystems. Finally, we review two important case studies on plasmonic microsystems, namely, the plasmonic NSOM and surface manipulation using slits and nanogratings.

### A. Electromagnetic theory

Electromagnetic wave interactions with metals can be fully described using the Maxwell's equations in a conventional macroscopic framework. In macroscopic frame, charges of singular nature and their coupled currents are approximated by referring to charge density $\rho$ and current density $J$. In differential form, Maxwell's equations can be written as

$$\nabla \times \vec{\mathrm{E}}(\mathbf{r},t) = -\frac{\partial \vec{\mathrm{B}}(\mathbf{r},t)}{\partial t}, \qquad (1)$$

$$\nabla \times \vec{\mathrm{H}}(\mathbf{r},t) = \frac{\partial \vec{\mathrm{D}}(\mathbf{r},t)}{\partial t} + \vec{\mathrm{J}}(\mathbf{r},t), \qquad (2)$$

$$\nabla \bullet \vec{\mathrm{D}}(\mathbf{r},t) = \rho(\mathbf{r},t), \qquad (3)$$

$$\nabla \bullet \vec{\mathrm{B}}(\mathbf{r},t) = 0, \qquad (4)$$

These equations couple the four macroscopic fields $D$ (the electric flux density), $E$ (the electric field), $H$ (the magnetic field), and $B$ (the magnetic flux density) with the exterior charge ($\rho$) and current ($J$) densities. In terms of macroscopic electromagnetics, the properties of a medium can be generally examined in terms of the polarization $P$ and magnetization $M$ according to:

$$\vec{\mathrm{D}}(\mathbf{r},t) = \varepsilon_0\, \vec{\mathrm{E}}(\mathbf{r},t) + \vec{\mathrm{P}}(\mathbf{r},t), \qquad (5)$$

$$\vec{\mathrm{H}}(\mathbf{r},t) = \mu_0^{-1}\, \vec{\mathrm{B}}(\mathbf{r},t) - \vec{\mathrm{M}}(\mathbf{r},t), \qquad (6)$$

Where $\varepsilon_0$ and $\mu_0$ are the permittivity and the permeability of the vacuum, respectively. For an isotropic, linear, and nonmagnetic approximation, the constitutive relations can be written as:

$$\vec{\mathrm{D}}(\mathbf{r},t) = \varepsilon_0 \varepsilon_r\, \vec{\mathrm{E}}(\mathbf{r},t), \qquad (7)$$

$$\vec{\mathrm{B}}(\mathbf{r},t) = \mu_0 \mu_r\, \vec{\mathrm{H}}(\mathbf{r},t), \qquad (8)$$

where $\varepsilon_r$ and $\mu_r$ are the relative permittivity and relative permeability, respectively. For a nonmagnetic medium we can flout the magnetic response, $\mu_r = 1$. The last important constitutive linear relationship is

$$\vec{\mathrm{J}}(\mathbf{r},t) = \sigma\, \vec{\mathrm{E}}(\mathbf{r},t) \qquad (9)$$

where $\sigma$ is conductivity. The wave equation is a useful description of many types of waves, from light and sound to water waves. To derive the wave equation, we take the curl of Eq. (1). Using Eq. (2), for a source-free region, the wave equation can be written as

$$\nabla^2\, \vec{\mathrm{E}}(\mathbf{r},t) = \mu\varepsilon \frac{\partial^2\, \vec{\mathrm{E}}(\mathbf{r},t)}{\partial t^2}, \qquad (10)$$

Here $\varepsilon$ and $\mu$ are the electric permittivity (also called absolute permittivity, $\varepsilon = \varepsilon_0 \varepsilon_r$) and magnetic permeability ( $\mu = \mu_0 \mu_r$ ). Eq. (10) This is actually three equations, which together comprise the $x$-, $y$- and $z$- vector components for the E field vector.

### B. Surface wave electromagnetics

Surface waves play a key role in studying properties of condensed matter at the interface between two media. Surface waves can be simply defined as waves propagating along the interface between two media and existing in both of them. Scientific interest in surface waves stems from their ability to confine the energy of wave to the close vicinity of the interface of two partnering materials. Any change in the composition of either partnering material in that vicinity could alter-even eliminate- the surface wave.

Zenneck in 1907 [11] initiated the concept of surface waves. He envisioned that the planar interface between air and ground supports the propagation of radiofrequency waves. The idea was later mathematically extended by Sommerfeld, and those surface waves have since become known as the Zenneck waves [12]. The concept of Zenneck waves were employed in the mid-20th century at the interface of a noble metal and a dielectric material in the visible range of the electromagnetic spectrum, which instigated the concept of surface plasmon polariton (SPP) waves. Since then, there has been significant advances in both theoretical and experimental investigations of SPPs [13, 14].

For the metal $\varepsilon_m$ at y< 0, and for the dielectric $\varepsilon_d$ at y> 0, the wave equation in Eq. (10) has to be solved separately in each region.

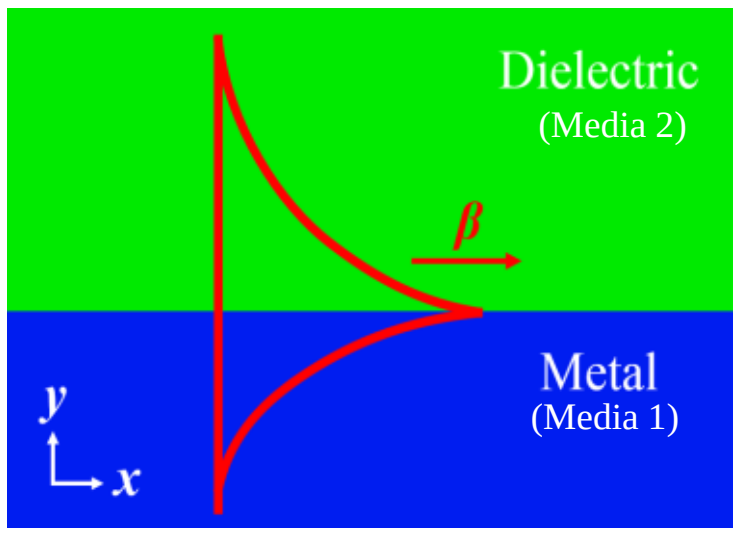

Fig. 2. Schematic of a surface wave propagating along a single metal-dielectric interface.

The solution of Maxwell's equations gives two sets of independent responses with different polarizations – TM (also known as p-polarized or vertical polarization) and TE (also known as s-polarized or horizontal polarization) modes. The TE mode has an E field that is perpendicular to the plane of incidence; TM mode has an E field that is parallel to the plane of incidence. Surface plasmon waves do not support a TE mode since the width of the propagation channel (waveguide) is much smaller than the exciting wavelength. Considering the following boundary conditions for the E field in media 1 (x-direction ($E_{x1}$) and y-direction ($E_{y1}$)), and media 2 (x-direction ($E_{x2}$) and y-direction ($E_{y2}$)),

$$E_{x1} - E_{x2} = 0, \tag{11}$$

$$\varepsilon_m E_{y1} - \varepsilon_d E_{y2} = 0, \tag{12}$$

These equations show that the parallel field component is continuous, whereas the perpendicular component is discontinuous. Solving Eq. (12) along with Eq. (3), the following relations for the wave vector can be obtained.

$$k_x^2 + k_{y_j}^2 = \varepsilon_j k^2 = \varepsilon_j (\frac{\omega}{c})^2 \rightarrow k_{y_j} = \sqrt{\varepsilon_j (\frac{\omega}{c})^2 - k_x^2} \qquad j = 1,2 \tag{13}$$

Eq. (13) exhibits the dispersion relation between the wave vector components and the angular frequency.

### C. *Dielectric constant of metals*

We start by deriving the dielectric constant of metals. One of the most straightforward but nevertheless valuable models to describe the response of a metallic particle exposed to an electromagnetic field is the Drude-Sommerfeld model

$$m_e \frac{\partial^2 r}{\partial t^2} + m_e \gamma_d \frac{\partial r}{\partial t} = eE_0 e^{-i\omega t}, \tag{14}$$

where $e$ and $m_e$ are the charge and effective mass of the free electrons, $r$ is the displacement of electron cloud due to the applied electric field, and $E_0$ and $\omega$ are the amplitude and frequency of the applied electric field. The damping term $\gamma_d$ is proportional to $\gamma_d = \upsilon_F / l$, where $\upsilon_F$ is the Fermi velocity and $l$ is the electron mean free path between scattering events. Eq. (14) can be solved by $r(t) = r_0 e^{-i\omega t}$, which leads to the well-known dielectric function of Drude form

$$\varepsilon_m(\omega) = \varepsilon_\infty - \frac{\omega_p^2}{\omega^2 + i\gamma_d \omega}, \qquad \omega_p = \sqrt{\frac{4\pi n_e e^2}{m_e}}, \tag{15}$$

where $\omega_p$ is the plasma frequency, and $\varepsilon_\infty$ expresses the ionic background in the metal (typically 3.7 for silver). If $\omega$ is greater than $\omega_p$, the corresponding refractive index ($n = \sqrt{\varepsilon_m}$) is a real number ; on the other hand, if $\omega$ is less than $\omega_p$, the refractive index is imaginary since $\varepsilon_m$ is negative.

### D. *Surface plasmon electromagnetics*

As shown in Fig. 2, the plane interface between a metal at $y$< 0, with a complex dielectric constant $\varepsilon_m$, in which the real part is negative (metals at terahertz (THz) region have negative real permittivity this is a critical criterion for SPP waves since in this situation the wave can actually penetrate inside the metal), and a dielectric at $y$> 0, with a positive dielectric constant is the most straightforward geometry that can support a SPP wave. As mentioned earlier, only the TM mode ($H_z$, $E_x$, and $E_y$) solutions are taken into account for SPP propagation. Propagating waves can be described as $\mathrm{E}(x, y, z) = \mathrm{E}(y)\mathrm{e}^{i\beta x}$ in which $\beta = k_x$ is called the propagation constant of the traveling wave. By taking into account Eq. (1) and Eq. (2) and identifying the propagation along the $x$-direction ($\partial / \partial x = i\beta$) and homogeneity in the $z$-direction ($\partial / \partial z = 0$) and ($\partial / \partial t = -i\omega$), the following equations can be derived to describe the TM mode

$$\vec{\mathrm{E}}_x = i \frac{1}{\omega \varepsilon_0 \varepsilon} \frac{\partial \vec{H}_z}{\partial y} \tag{16}$$

$$\vec{\mathrm{E}}_y = \frac{\beta}{\omega \varepsilon_0 \varepsilon} \vec{\mathrm{H}}_z \tag{17}$$

and the wave equation for the $H_z$ component as

$$\frac{\partial^2 \mathrm{H}_z}{\partial y^2} + (k_0^2 \varepsilon - \beta^2) \mathrm{H}_z = 0 \tag{18}$$

For the half-space in Fig. 2, the TM wave can be written for upper and lower halves as

$$\mathrm{H}_z(\mathrm{y}) = \mathrm{A}_2\, e^{i\beta x} e^{-k_2 y} \tag{19}$$

$$E_x(y) = -i A_2 \frac{k_2}{\omega\varepsilon_0\varepsilon_d} e^{i\beta x} e^{-k_2 y} \tag{20}$$

$$E_y(y) = A_2 \frac{\beta}{\omega\varepsilon_0\varepsilon_d} e^{i\beta x} e^{-k_2 y} \tag{21}$$

for $y > 0$, and

$$H_z(y) = A_1 e^{i\beta x} e^{k_1 y} \tag{22}$$

$$E_x(y) = i A_1 \frac{k_1}{\omega\varepsilon_0\varepsilon_m} e^{i\beta x} e^{k_1 y} \tag{23}$$

$$E_y(y) = A_1 \frac{\beta}{\omega\varepsilon_0\varepsilon_m} e^{i\beta x} e^{k_1 y} \tag{24}$$

for $y < 0$. Due to boundary conditions and the wave equation for $H_y$, we have

$$A_1 = A_2, \quad \frac{k_2}{k_1} = -\frac{\varepsilon_d}{\varepsilon_m}, \quad \begin{matrix} k_1^2 = \beta^2 - k_0^2\varepsilon_m \\ k_2^2 = \beta^2 - k_0^2\varepsilon_d \end{matrix} \tag{25}$$

Finally, the dispersion relation of SPPs propagating at the interface between the interface of two half spaces can be obtained as

$$\beta = k_0\sqrt{\frac{\varepsilon_m\varepsilon_d}{\varepsilon_m + \varepsilon_d}} \tag{26}$$

Fig. 3 presents the dispersion of surface plasmons as a function of wavevector. It can be seen from Fig. 3 that for a given frequency, a free-space photon has less momentum than an SPP since they don't intersect. On the other hand, coupling medium such as a prism can match the photon momentum due to higher refractive index compared to free-space which shifts the prism-coupled dispersion curve so that the curves intersect. SPPs coupling with a prism light with relative permittivity of 1.5 with silver is presented in Fig. 3.

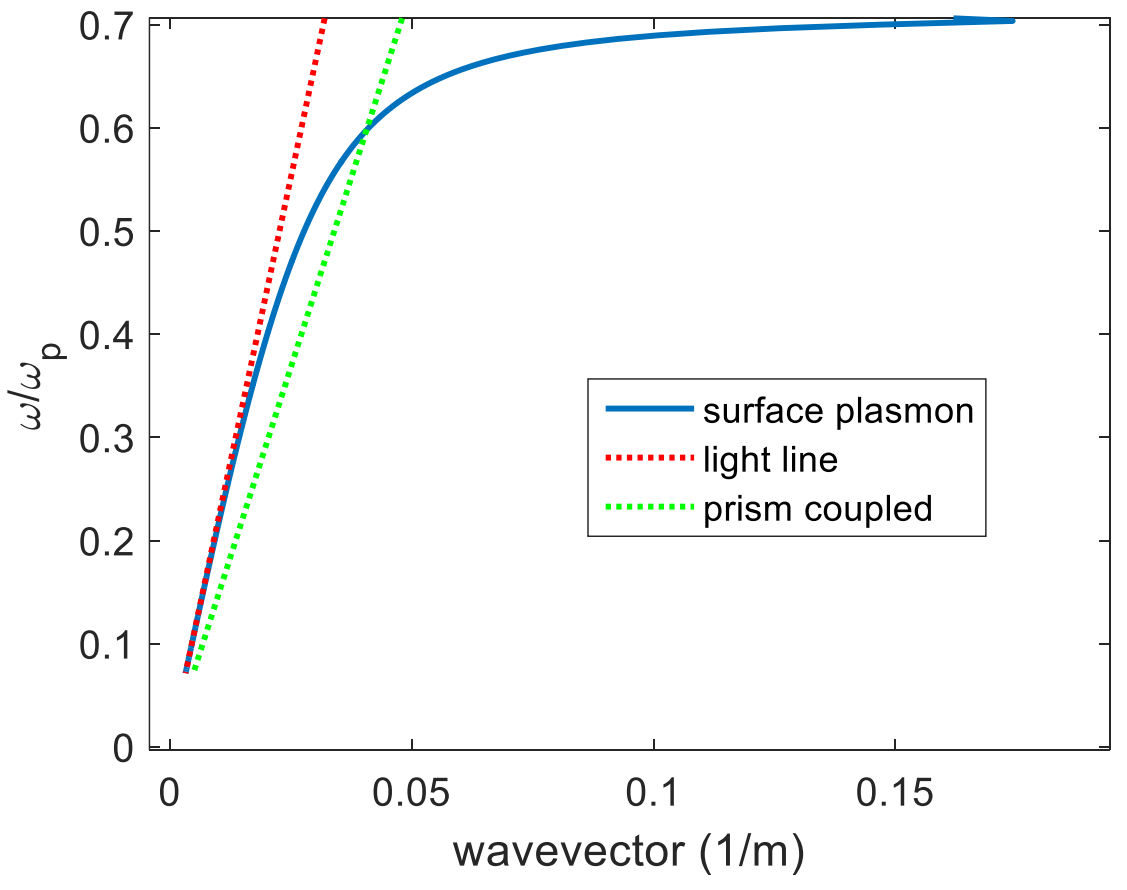

Fig. 3. Dispersion relation of SPPs at the interface of a metal and a dielectric.

## E. *Transmission line theory (TLT)*

Transmission line theory (TLT) is a fully analytical, fast, and reliable methodology for analyzing wave propagation in metallic and dielectric structures. It has been extensively exploited to investigate the wave propagation in plasmonic structures. In this section, we briefly review the literature on TLT concerning plasmonic microsystems and biosensing applications.

Considering a metal-dielectric-metal (MDM) waveguide in Fig. 4 consisting of a dielectric layer of thickness *h* surrounded by two metallic layers. For *h* much smaller than the wavelength, only the fundamental transverse magnetic (TM) waveguide mode can propagate along the waveguide.

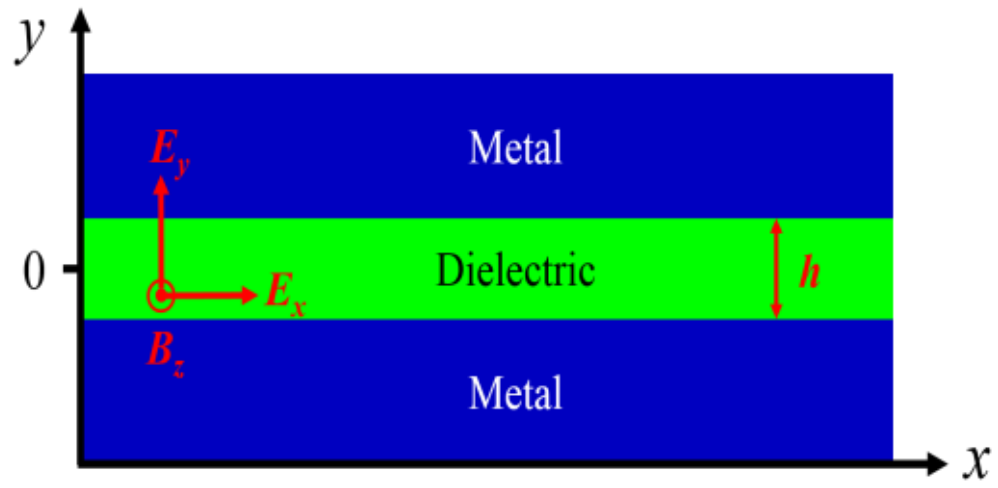

Fig. 4. Surface plasmon MDM waveguide with dielectric layer of thickness *h*. Only the TM mode can propagate inside the waveguide for *h* much smaller than the wavelength.

The effective refractive index of an MDM waveguide can be calculated via [15]

$$n_{eff} = \frac{\beta}{k} = \frac{\lambda}{\lambda_{MDM}} + i\frac{\lambda}{4\pi L_{SPP}} \tag{27}$$

where $k = 2\pi/\lambda$ is the wavenumber. The real part of Eq. (27) describes the guided wavelength ($\lambda_{MDM}$) and the imaginary part determines the propagation length ($L_{SPP}$). The complex-valued propagation constant ($\beta$) can be obtained via the following dispersion relation for the TM-SPP modes [16]

$$\tanh\left(\frac{ik_1 h}{2}\right) = \left(\frac{\varepsilon_2 k_1}{\varepsilon_1 k_2}\right)^{\pm 1} \tag{28}$$

where $\pm$ sign presents the symmetric and antisymmetric resonance modes, and $k_j = (\varepsilon_j \mathrm{k}^2 - \beta^2)^{1/2}, (\mathrm{j} = 1, 2)$ where $\varepsilon_1$ and $\varepsilon_2$ are the relative permittivity of the dielectric and metal, respectively. When the width of the MDM waveguide is not too small, Eq. (28) can be further approximated as

$$\beta(\mathrm{h}) = k_0\sqrt{\varepsilon_1 - 2\varepsilon_1\frac{\sqrt{\varepsilon_1 - \varepsilon_2}}{k_0 h \varepsilon_2}} \tag{29}$$

in which $k_0$ is the optical wave vector in vacuum.

Our first example considers a recent work by Li et al. [17] in which they have proposed a THz MDM waveguide sensor with an embedded microfluidic channel. The proposed structure is suitable for sensing the refractive index variations in liquid. The proposed THz waveguide with a two layer-stub structure is conceptually outlined in Fig. 5a. The transmission spectrum is analytically described using the TLT method and compared with the numerical results obtained using the finite-difference time domain (FDTD) method. Initially, the waveguide of width *d* is substituted by a transmission line of characteristic impedance of

$$Z(d)=\frac{\beta(d)d\eta}{k_0\varepsilon_d} \qquad (30)$$

where $\varepsilon_d$ and $\eta$ are the relative permittivity of dielectric and wave impedance in the dielectric, respectively, and $\beta$ has been obtained using Eq. (29). The impedance of medium can be related to the impedance of free space $\eta_0$ using $\eta=\eta_0/n$ (inversely proportional to refractive index). The TLT equivalent network of the biosensor presentation is presented in Fig. 5c. The effective impedance corresponding to the two layer-stub fraction of transmission can be derived by

$$Z_{stub}=Z_d\frac{Z_L'+jZ_d\tan(\beta_d\,\mathrm{h}_1)}{Z_d+jZ_L'\tan(\beta_d\,\mathrm{h}_1)} \qquad (31)$$

in which $\beta_d$ is the propagation constant of surface plasmons in the spacing dielectric, and $Z_L'$ is the effective impedance for the liquid microfluidic sample and is calculated via

$$Z_L'=Z_s\frac{Z_L+jZ_s\tan(\beta_s\,\mathrm{h}_2)}{Z_s+jZ_L\tan(\beta_s\,\mathrm{h}_2)} \qquad (32)$$

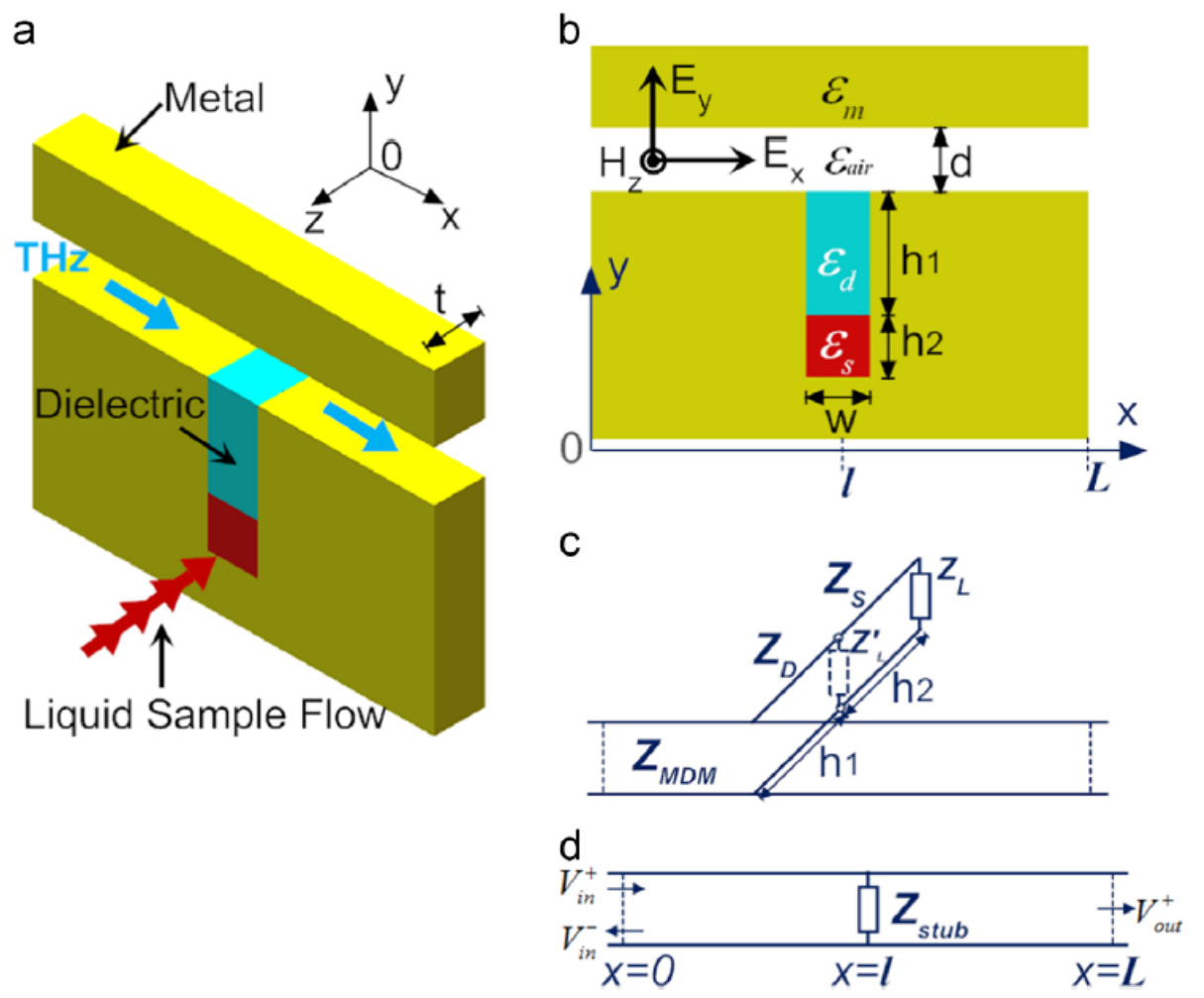

Fig. 5. The proposed structure of a THz MDM waveguide with a two stacked sensing stubs including a dielectric spacer and a sample of liquid biomaterial layer (a) 3-D Schematic (b) 2-D Schematic (c-d) the equivalent transmission-line representation, and its equivalent circuit model [17].

where $Z_d$ and $Z_s$ are the characteristic impedances corresponding to $\varepsilon_d$ and $\varepsilon_s$ (sample permittivity), respectively and $\beta_S$ is the SPP propagation constant in the microfluidic sample [18]. Finally, the transmission properties of the MDM waveguide calculated using the following formula

$$\mathrm{T}=\left|1+\frac{Z_{stub}}{2Z_{air}}\right|^{-2}\exp\left(-\frac{L}{L_{SPP}}\right) \qquad (33)$$

Eq. (33) can be further extended to model wave propagation in two and multiple stubs [15].

The second example includes a slit-grating analysis using the TLT method. The transmission properties of the plasmonic nanograting structures are usually coupled with sharp resonances over a narrow bandwidth [19]. This methodology has been used to enhance on-chip quantum dot cellular imaging resolution. In this section, theoretical analysis using the TLT method that was used by the authors to analyze the scattering from this structure has been reviewed. The details of the applications of the proposed technique will be reviewed in the applications section.

Initially, the proposed geometry is partitioned into small grids of transmission line (TL) unit-cells, each excited with a different phase $\phi=p\cdot k_0\sin\theta$ for a given incidence angle $\theta$. Next, the TLT method is applied for each unit cell. The surrounding medium is approximated via wavenumber $\beta_0=k_0\cos\theta$ and characteristic impedance per unit length $Z_0=p(\mu_0/\varepsilon_0)^{1/2}\cos\theta$. Beginning from the TM-mode dispersion in Eq. (28) and following the same procedure mentioned above, the reflection coefficient at the entrance of the slits can be obtained as

$$\mathrm{r_H}=\frac{iZ_\mathrm{S}\left(Z_\mathrm{T}-Z_0\right)+\left(Z_\mathrm{S}^2-Z_\mathrm{T}Z_0\right)\tan\beta_\mathrm{S}t_\mathrm{M}}{iZ_\mathrm{S}\left(Z_\mathrm{T}+Z_0\right)+\left(Z_\mathrm{S}^2+Z_\mathrm{T}Z_0\right)\tan\beta_\mathrm{S}t_\mathrm{M}}, \qquad (34)$$

in which the input impedances per unit length at the series junctions are

$$Z_\mathrm{T}=Z_\mathrm{D}\frac{Z_\mathrm{L}^+-iZ_\mathrm{D}\tan\left(\beta_\mathrm{D}l\right)}{Z_\mathrm{D}-iZ_\mathrm{L}^+\tan\left(\beta_\mathrm{D}l\right)}+Z_\mathrm{D}\frac{Z_\mathrm{L}^--iZ_\mathrm{D}\tan\left(\beta_\mathrm{D}l\right)}{Z_\mathrm{D}-iZ_\mathrm{L}^-\tan\left(\beta_\mathrm{D}l\right)} + Z_\mathrm{S}\frac{Z_0-iZ_\mathrm{S}\tan\left(\beta_\mathrm{S}t_\mathrm{M}\right)}{Z_\mathrm{S}-iZ_0\tan\left(\beta_\mathrm{S}t_\mathrm{M}\right)}. \qquad (35)$$

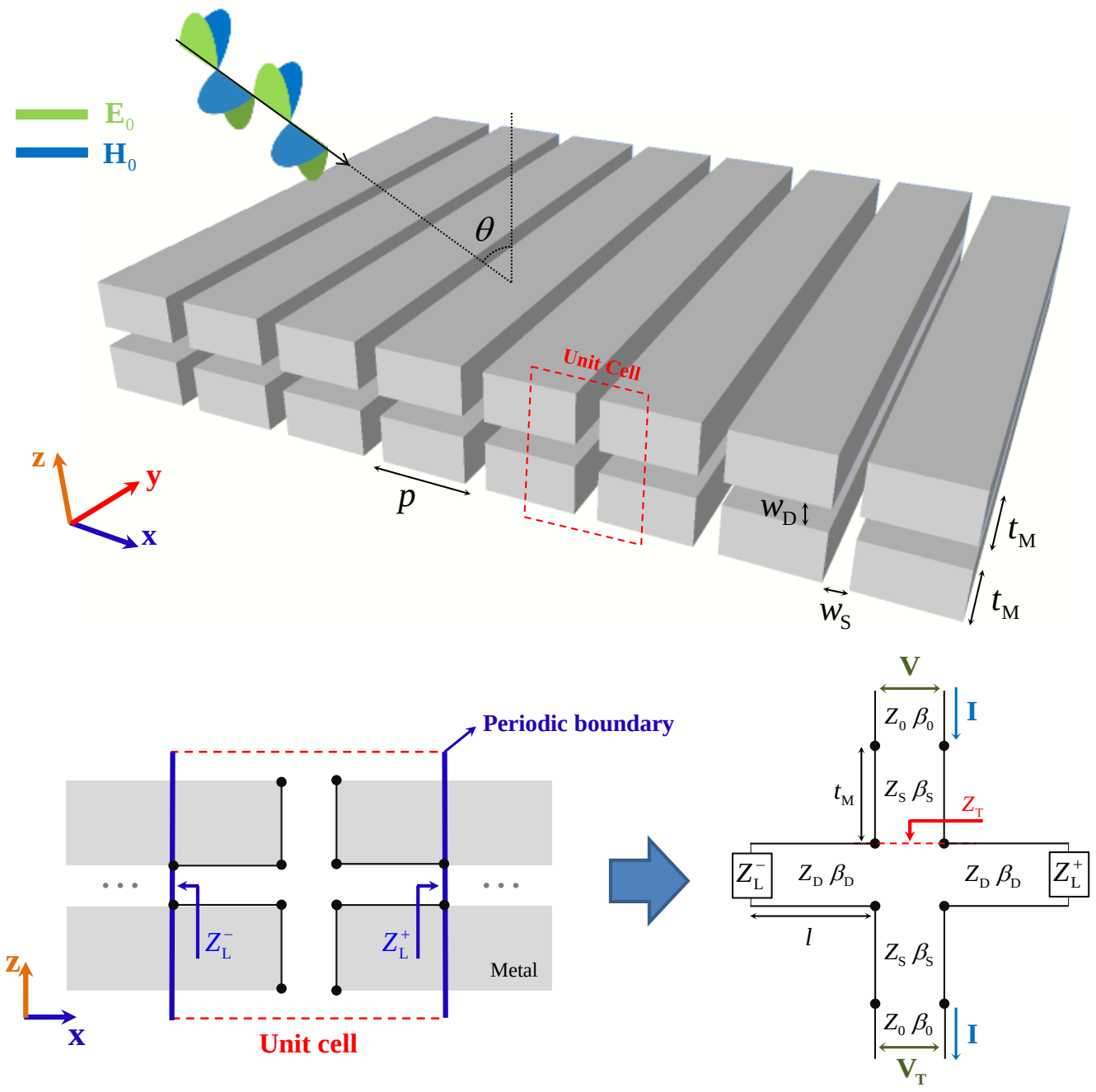

Fig. 6. A plasmonic nanograting device forming a loaded waveguide (top) partitioned structure into a series of unit cells. $\theta$ is the incident angle (bottom left) the corresponding TLT model. $p$ is the period, $l=\left(p-w_\mathrm{D}\right)/2$ is the stub length, $t_\mathrm{M}$ is the metal thickness, and, $w_\mathrm{D}$ and $w_\mathrm{S}$ are the widths of dielectric layer and slit, respectively. The equivalent load impedance $Z_\mathrm{L}^{\pm}$ can be readily calculated by evaluating the ratio for forward and backward magnetic-field waves at the periodic boundary [19].

It has been shown that the proposed structure can support sharp bandgap responses which can be used for spectral manipulation and effective suppression of light transmission.

### *F. Plasmonic NSOM tip theory*

Near-field scanning optical microscopy (NSOM) is a scanning probe based optical microscopy technique that can provide high-resolution optical imaging with sub-diffraction limit resolution by exploiting the properties of evanescent waves [20-22]. NSOM plasmonic tips can be considered as one the most interesting applications of plasmonic microsystems in which nano-scale plasmonic slits and grooves have been used to focus optical energy at the apex of a probe. The applications of plasmonic NSOM probes in near-field imaging will be reviewed in the applications section.

The simplest plasmonic geometry that can be mounted on a probe is a single-slit coupler in Fig. 7 [23]. The goal is to control and optimize the width of a slit in order to generate unidirectionally propagating SPPs in one direction and perfect reflections of the waves in all other directions. This condition will focus and guide the surface plasmon energy to the apex of the probe.

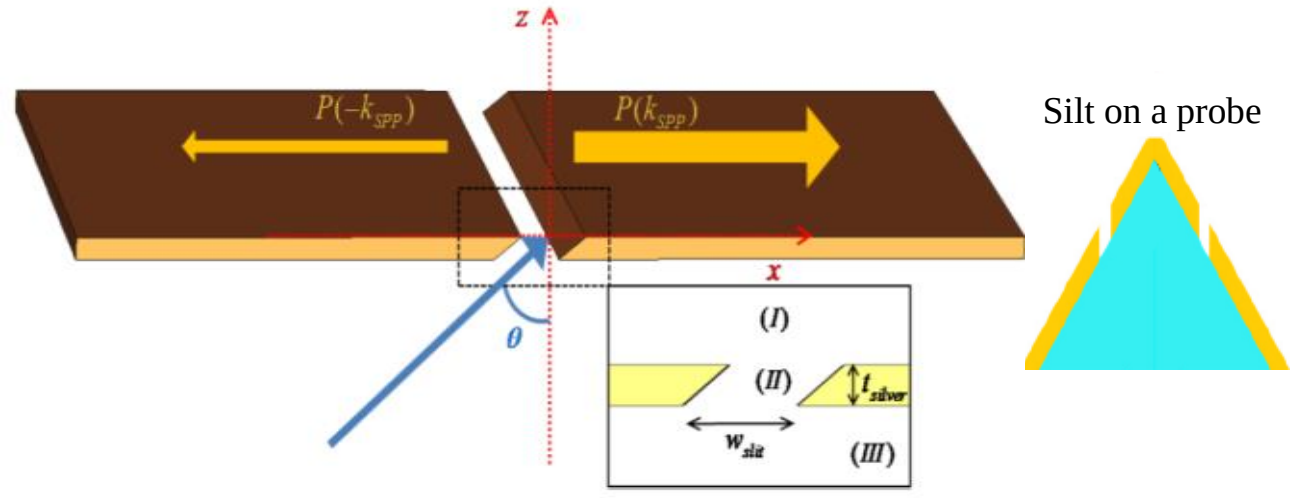

Fig. 7. A metal slit to excite unidirectional surface plasmons. The blue arrow indicates the direction of the wavevector of TM polarized plane wave incident on the backside of the thin metal with the incident angle Ө. $t_{slit}$ is the thickness of the metal film. (Inset shows how the silt can be mounted on a NSOM probe).

The angular spectrum of the slit can be derived as

$$A(k_x, z) = \int_{-\infty}^{\infty} u(x,z)\Pi(x,z)\exp[jk_x x]\,dx \qquad (36)$$

in which $\Pi(x,z)$ describes the slit geometry and $u(x,z)$ is the amplitude for a monochromatic wave in phasor form $U(x,z,t) = u(x,z)e^{j\omega t}$. For effective surface plasmon generation, the parameters that maximize the value of $A(k_{spp}, z = t_{slit})$ will be the optimized geometric parameters.

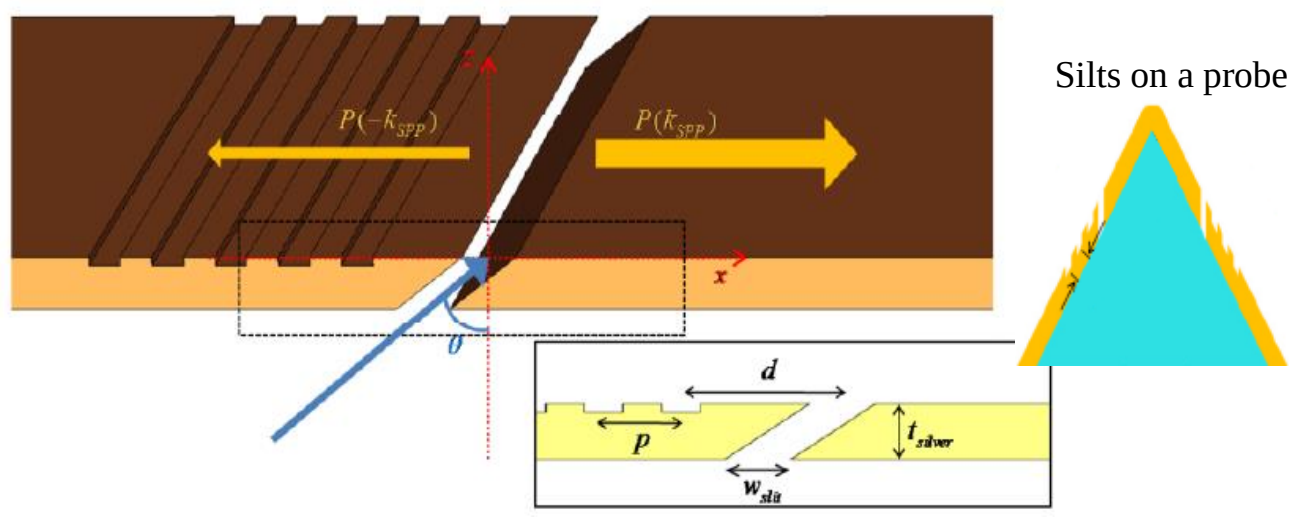

Fig. 8. Schematic drawing of plasmonic Nanograting on the left-had side of the slit in Fig. 7. The Nanograting structure is used to direct SPPs to the right-hand side. (Inset shows how the silt can be mounted on a NSOM probe).

Slit-grating and groove geometry in Fig. 8 can be used to generate more strongly confined unidirectional SPPs by using proper geometrical parameters. The slit nanograting on the left-hand side of the groove generates standing waves in harmony with the wavelength of surface plasmons inside the groove. A nanograting structure with $k_{spp} = n\pi / 2d$, where $n$ is the grating index of diffraction, is exploited to increase the directivity of SPPs by attaining destructive interferences on the left wall of the slit. At the exit point of the slit, the directivity of the surface plasmon excitations can be determined using

$$Directivity = \frac{\left|H_y^{right}\right|}{\left|H_y^{left}\right|} \qquad (37)$$

where $H_y^{right}$ is the y-directed magnetic field component determined at the right-side of the slit and $H_y^{left}$ is the magnetic field obtained at the left-side of the slit.

### *G. Plasmonic nanograting theory*

Since the original report of grating diffraction by Hopkinson and Rittenhouse [24], the interaction of light with periodic and nanograting structures have become a fascinating subject for understanding the nature of light-matter interaction. The wave diffraction offers the ability to control the spatial and spectral orders of light, which can be exploited in many notable applications such as color filters and fluorescence spectroscopy. Maisonneuve et al. [25] have exploited a nanograting structure as a coupling medium to match the photon and SP wavevectors. This idea was later extended to design a phase sensitive sensor on a plasmonic nanograting.

Zhang's group in [26] has exhibited a unidirectional optical beaming obtained via an array of sub-wavelength nano-slits on a metallic surface. The proposed geometry contains a subwavelength size slit with an array of periodic gratings (bumps) on its left-hand side which are patterned on a 250 nm Ag film as shown in Fig. 9.

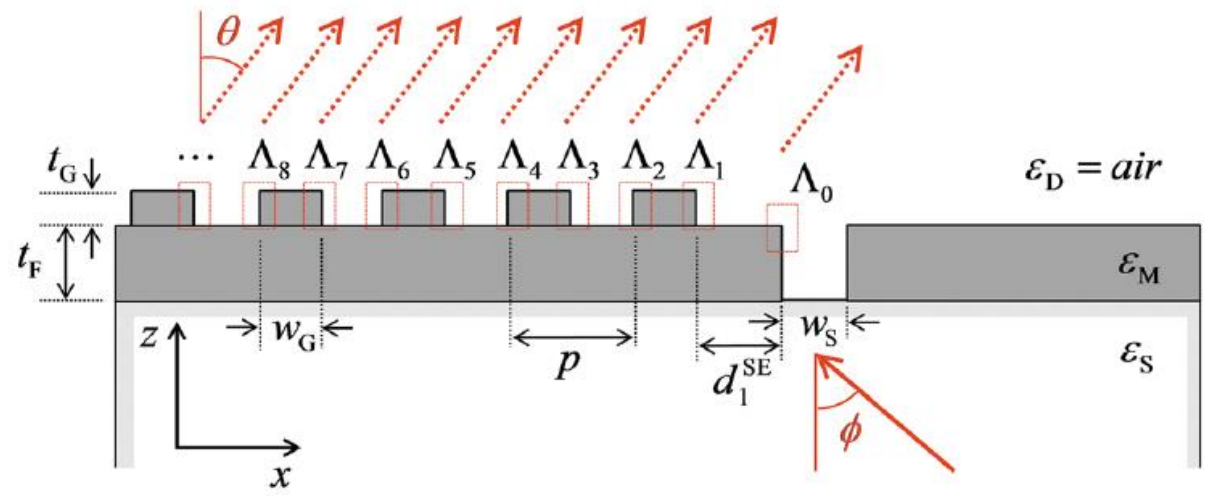

Fig. 9. Sub-wavelength slit with a single-sided array of periodic corrugations [26].

The nano-slit is used to couple the TM polarized incident light into SPPs. The scattered wave from each nanograting can be obtained using the wave equation in Eq. (10). The scattered wave is due to the interaction of surface plasmon waves and the nanograting. The scattered wave $E_i^S$ from each grating $\Lambda_i$ was calculated using the superposition of perturbed $E_i^P$ and unperturbed $E_i^U$ fields.

$$\begin{aligned} E_i^S(x,z) &= E_i^P(x,z) - E_i^U(x,z) \\ &= A_i \iint k_0^2[\varepsilon_P(x',z') - \varepsilon_U(z')] \\ &\times \bar{G}(x-x',z,z')\, E_i^P(x',z')\, d\Lambda_i' \end{aligned} \tag{38}$$

in which $\bar{G}$ is the Green's dyadic. Considering the phasor form, Eq. (38) can be written as

$$\begin{aligned} E_i^S &= \iint k_0^2[\varepsilon_P(x',z') - \varepsilon_U(z')] E_i^U(x',z')\, d\Lambda_i' \\ &\times \exp(-jk_0\,\rho + j\psi_0 - jk_{SP} d_i^{SE}) \end{aligned} \tag{39}$$

where $\varepsilon_P$ and $\varepsilon_U$ are the perturbed and unperturbed dielectric constants of the nanograting corresponding to $E_i^P$ and $E_i^U$ [26]. A phase matching condition has been used to obtain a collimated directional beam using

$$k_{SP} d_i^{SE} = \begin{cases} d_i^{SE}\, k_0 \sin\theta + 2\pi m_i,\, i = odd \\ d_i^{SE}\, k_0 \sin\theta + 2\pi m_i - \pi,\, i = even \end{cases}, \tag{40}$$

where $m_i$ is the diffraction order at the $i$th edge. The presented results demonstrate the applicability of this approach to design efficient directional plasmonic beaming using nanograting arrays.

In this section, we have reviewed the fundamental EM theories that shape the basis for the study of surface plasmonic waves. Starting with the Maxwell's equations, we have derived the dispersion relation for surface plasmons. The solution of wave equation supports two independent modes, TE and TM modes. We showed that the TLT method can be used to analytically analyze wave propagation in many plasmonic microsystem devices. Although the aforementioned methods can be applied to investigate EM scattering from a variety of plasmonic structures, numerical methods are more desirable for complex plasmonic microsystem devices. Novel, fast, and versatile numerical and analytical methods are currently under development for plasmonic microsystem structures [14, 27].

## III. Fabrication

The practicability of plasmonic microsystems depends upon the ability to fabricate these devices in the lab. Numerous nanofabrication techniques exist, each of which offers unique advantages and disadvantages in terms of performance, materials, applications, and geometry of the desired structure. It is not our goal in this paper to provide a comprehensive text on nanofabrication methodologies. Rather, this section aims to review the body of literature with plasmonic patterns on micro-scale devices and to highlight recent advances in plasmonic fabrication and nanolithography.

Because most materials depict properties and features considerably different from their bulk form in dimensions below 100-nm, the 100-nm scale has been settled as the threshold between nanotechnology and conventional microscale technologies. Here, we review the fabrication of sub-100-nm resolution plasmonic patterns on micro scale devices. Readers seeking a more detailed discussion regarding general nanofabrication techniques are referred to the following canonical books [28-31].

In this section, we classify plasmonic microsystem fabrication techniques by the resolution achievable as well as the associated physical principles. We also introduce a variety of methodologies that are currently under development in the laboratory.

### A. Photolithography

Photolithography, or "optical lithography", is the most reliable and least expensive lithography technology currently available for industrial microfabrication. Although it was initially designed for the microelectronics industry, photolithography has proven to be a powerful tool in plasmonic microsystems. For example, fabrication of 100 nm line patterns and 70 nm isolated lines have been realized using an intermediate hard mask material such as silicon oxide or silicon oxynitride [32]. Patterning with sub 100-nm resolution using a single layer of deep-UV photoresist (175 nm thick) has also been demonstrated (by using Sandia's 10x-Microstepper Extreme ultraviolet (EUV) imaging system) [33]. At even shorter X-ray wavelengths of light, photolithography was used to fabricate structures as small as 50 nm and at a limit even sub-30 nm [34-36]. Researchers at ENEA Frascati Research Center in 2008 reported the fabrication of structures with less than 90 nm using EUV at 14.4 nm [37, 38]. EUV lithography is now at the technological forefront in the challenge to miniaturize electrical components for semiconductor industry. Recently in 2015, researchers at the Lawrence Berkeley National Laboratory [39] engineered patterns at dimensions less than 20 nm with a low-cost photoresist. Furthermore, photolithography has been recently applied in the fabrication of subwavelength structures that operate at THz frequency.

The pioneering work regarding photolithography for plasmonic and metamaterial devices has been reported by Yen et al. [40] in which they have created an array of nonmagnetic and conducting split-ring resonators (SRRs), as shown in Fig. 10.

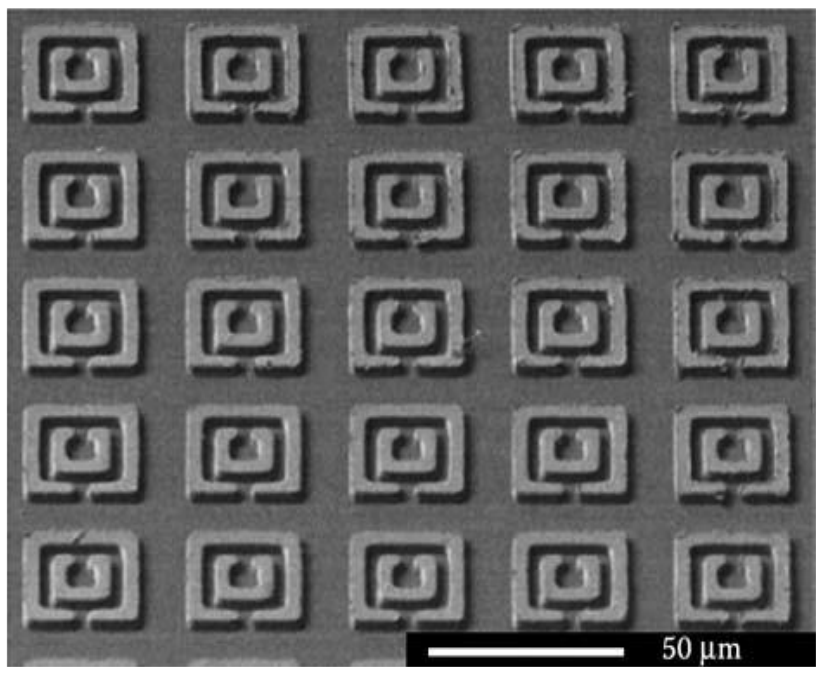

Fig. 10. Realized magnetic metamaterial structure using photolithography. Figure taken from [40].

Their metamaterial structure was fabricated via a special photolithographic technique termed “photo-proliferated process (PPP)” on a 400-μm-thick quartz substrate. The process is diagrammed in Fig. 11.

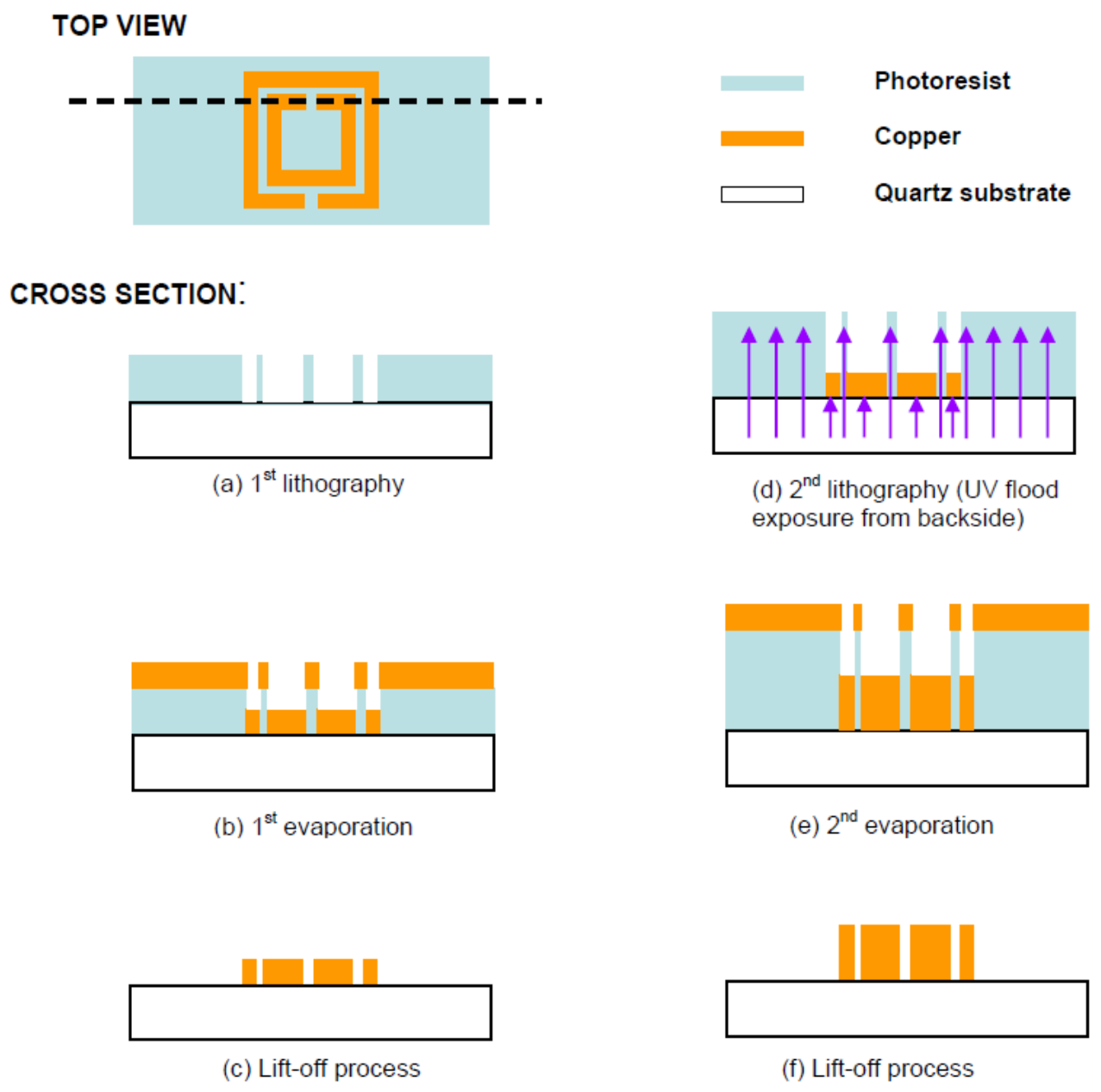

Fig. 11. Fabricated magnetic metamaterial structure using photolithography [40].

Initially, a negative photoresist (PR) with the thickness of 5 μm is spun onto the polished quartz substrate. Next, the designed SRR pattern on the photomask is transferred by a contact-mode lithographic process. After the pattern transfer, 100 nm thick chromium (Cr) and 1 μm thick copper (Cu) layers are evaporated to fill inside the SPR patterns using an e-beam evaporator. Lastly, an acetone solvent rinse in an ultrasonic bath is employed for the lift-off process. The result of the first step is a conductive, SRR shaped non-magnetic layer, patterned on the transparent quartz substrate.

In the second phase of lithography, another layer of PR is spun on the SRR mold. Eventually, a second copper evaporation and lift-off process are repeated to heighten the thickness of the conductive SRR structures.

Martin et al. in 2010 [41] have used photolithography along with shadow evaporation to fabricate sub-10 nm metallic (including gold) gap sizes for plasmonic applications as shown in Fig. 12. In their proposed two-stage method, first an EUV interference lithography was exploited to pattern a 1D line array on the glass or silicon substrate. A 13.5nm wavelength coherent beam was used to transfer two identical gratings on a PR. The diffracted beams caused by the gratings overlap to form high resolution patterns with dimensions below 10 nm half pitch. Secondly, glancing angle deposition (GLAD) was used to deposit metal layers on the PR.

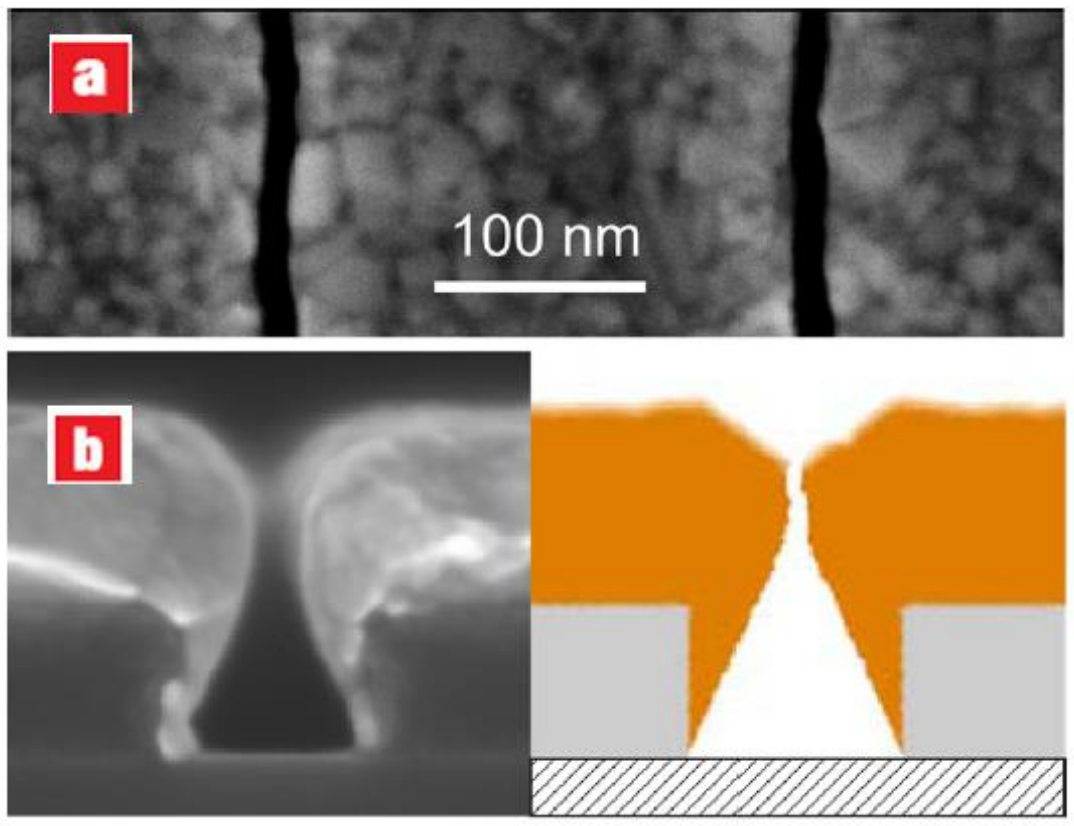

Fig.12. (a) SEM image of the top and (b) cross sectional view of a sliced gold nanogap array compared with ballistic simulation results. The angle of evaporation is 60 degrees from the surface normal. The resulting gap size is approximately 13 nm. Figure adapted from [41].

## B. *Nanofabrication using charged beams*

Nanofabrication by charged beams, including electron beam lithography (EBL) and focused ion beam lithography (FIB), is based on carrying and shooting energized particles onto a substrate material in order to perform structuring either by exposure of energy-sensitive polymer materials or by removing material directly. The major advantage of using charged beams is that they can be focused into extremely small regions. Other advantages of charged beams lithography over the photolithography techniques involve very high resolution and versatile pattern formation [42] which are critical for plasmonic microsystems [43].

El-Sayed et al. at Georgia Tech employed EBL to fabricate pairs of gold nanoparticles with altering interparticle distance, in order to study the effect of plasmon waves on the phonon oscillation [44]. The gold nanoparticle pairs were drawn on a quartz glass slide using a JEOL JBX-9300FS 100 kV EBL system. A 65-nm layer of PMMA polymer (an electron-sensitive resist) was spin-coated on the transparent quartz slide and cured at 180 °C for 3 min. Next, a 10-nm thick gold layer was mounted on the substrate using a thermal evaporator to make the substrate conductive. The structure is shown in Fig. 13.

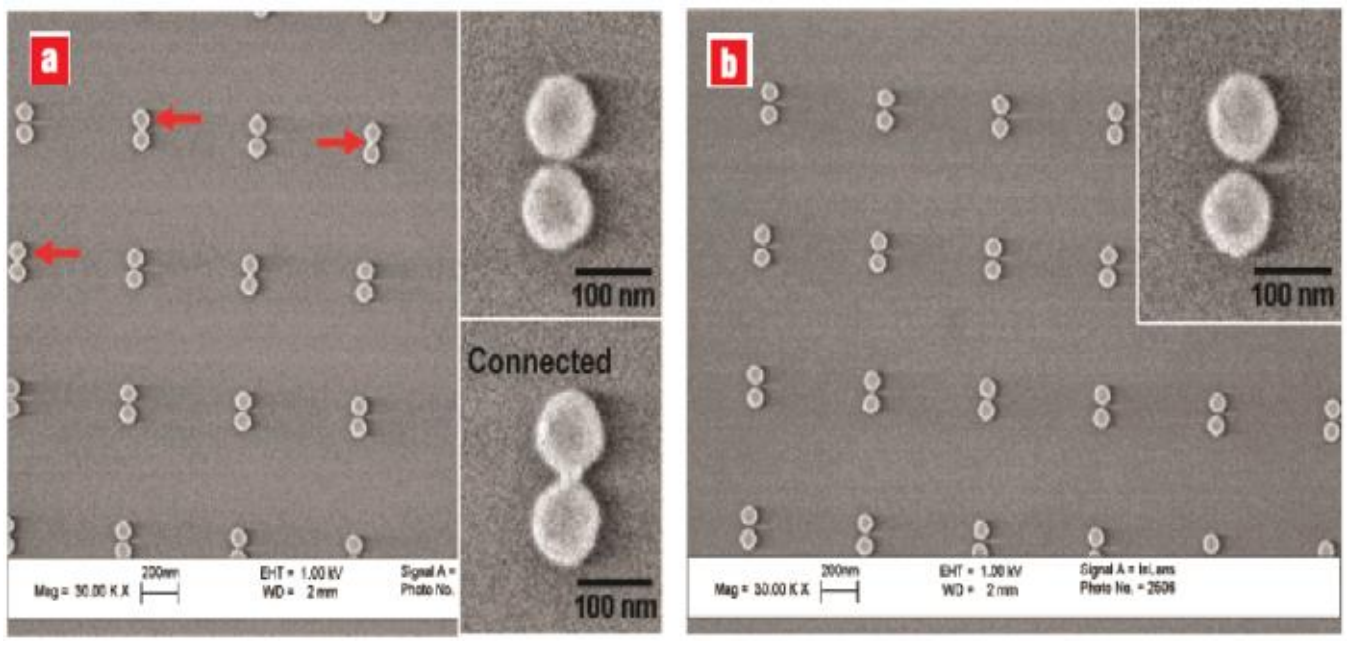

Fig. 13. SEM images of several nanodisc pairs having interparticle gaps of (a) 2 nm, (b) 7 nm, the insets presents the amplified images of the nanodisc pair [44].

Giessen's laboratory at the Universität Stuttgart produced a metamaterial structure consisting of an array of a stack of two identical SRRs [45] fabricated using the EBL and layer-by-layer stacking method. In the first step, 250-nm thick gold nanorod structures are fabricated in positive resist by EBL. Next a 100-nm-thick solidifiable photo polymer spacer is coated on the first layer. Subsequently, a second SRR structure was fabricated on the sample using gold film evaporation and EBL. The total area of the fabricated structures was 200 × 200 μm. In addition to bi-layer structures, multilayer structures have been also fabricated using EBL [46]. Some modification of EBL like scanning transmission electron microscope (STEM) are capable of 2 nm [47] and even 0.7 nm patterning using electron beam-induced deposition (EBID) [48]. This fabrication method can be used to fabricate multilayer plasmonic structures such as plasmonic Yagi–Uda antenna and perfect plasmonic absorbers [49].

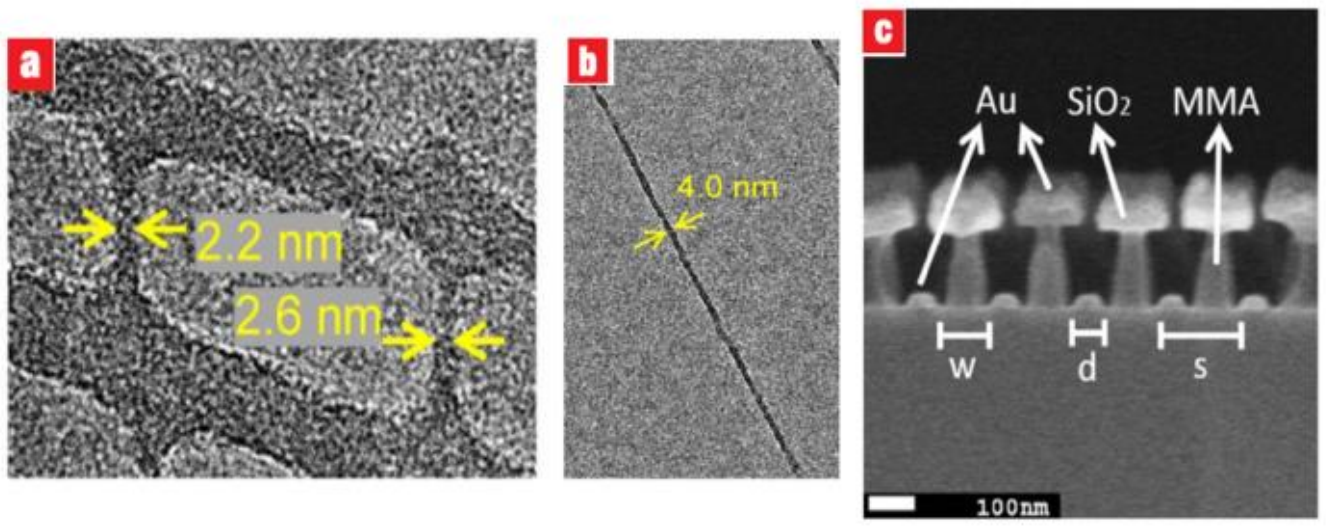

Fig. 14. (a) Gap sizes about 3 nm size obtained by EBL in [47], (b) An isolated feature with average line width of 4±0.8 nm [47], SEM cross-sectional micrograph of a tri-layer lift-off process after Au evaporation [46].

Zhang's group used EBL to experimentally demonstrate tunable plasmonic radiation from a periodic array of plasmonic nano-scatterers [26, 50]. The proposed structure can be tailored to convert surface plasmon polaritons into directional, high gain leaky modes. The schematic of the proposed tunable directional optical antenna is shown in Fig. 15.

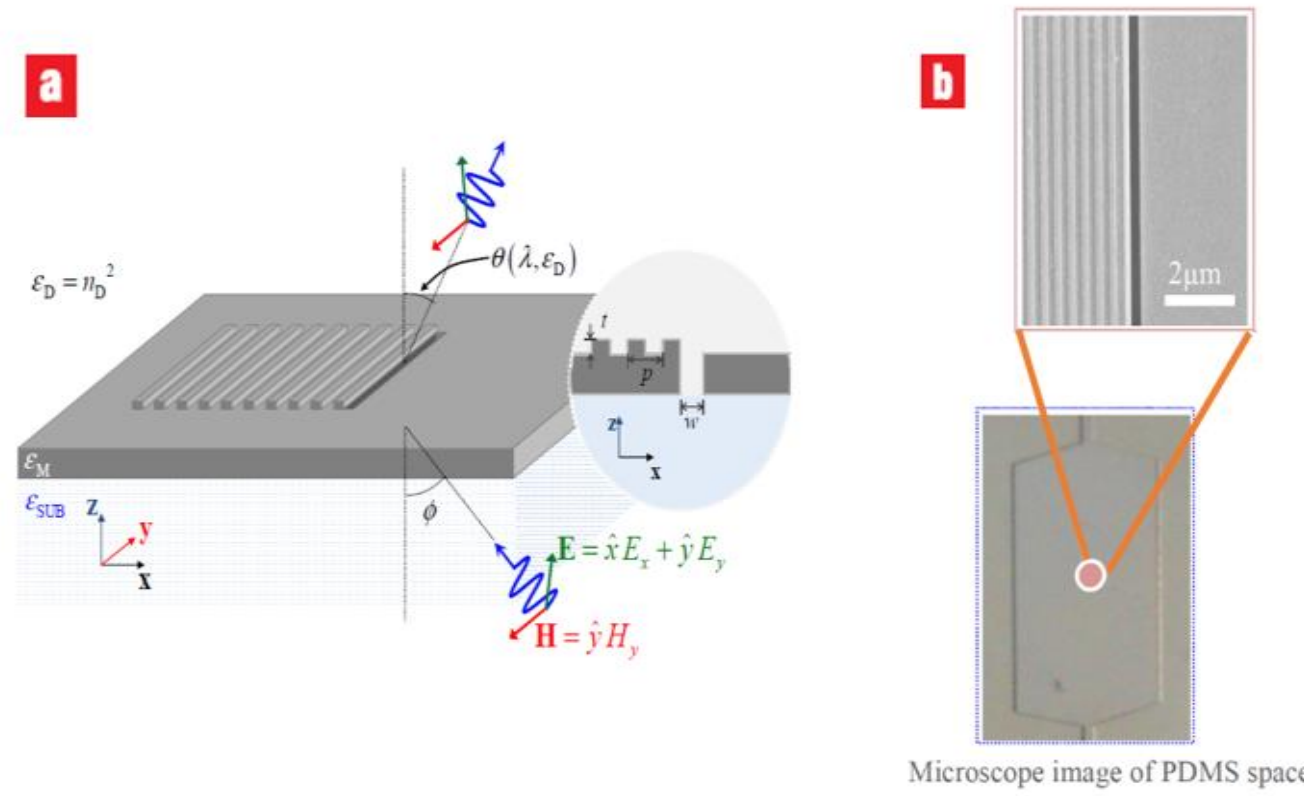

Fig. 15. (a) Schematic of the proposed tunable directional optical antenna. The subwavelength slit is supported with an array of periodic nanograting from the left-hand surface to confine and guide the confined energy of SPPs into one specific direction. $\varepsilon_D$, $\varepsilon_M$, and $\varepsilon_{SUM}$ indicate the relative permittivity of surrounding medium, metal, and supporting substrate (glass slide), respectively. Directive radiation at an explicit angle θ can be reached by an appropriate choice of surrounding medium and operating wavelength λ. The directivity of the structure can be further improved by optimizing the illumination angle φ. (b) SEM of the fabricated device [50].

As mentioned in the theory section, NSOM is one the most fascinating applications of plasmonic microsystems. For NSOM, probe-based nanoantennas are the key technical units. Taminiau et al. used a focused Ga+ ion beam [51] to create a precise stretched probe with a 50 nm width and 20 nm radius of curvature. The device sports a 100-nm aperture on a flat end face. Fig. 16 shows the scanning electron microscopy (SEM) images of the resulting probe-based antenna [52].

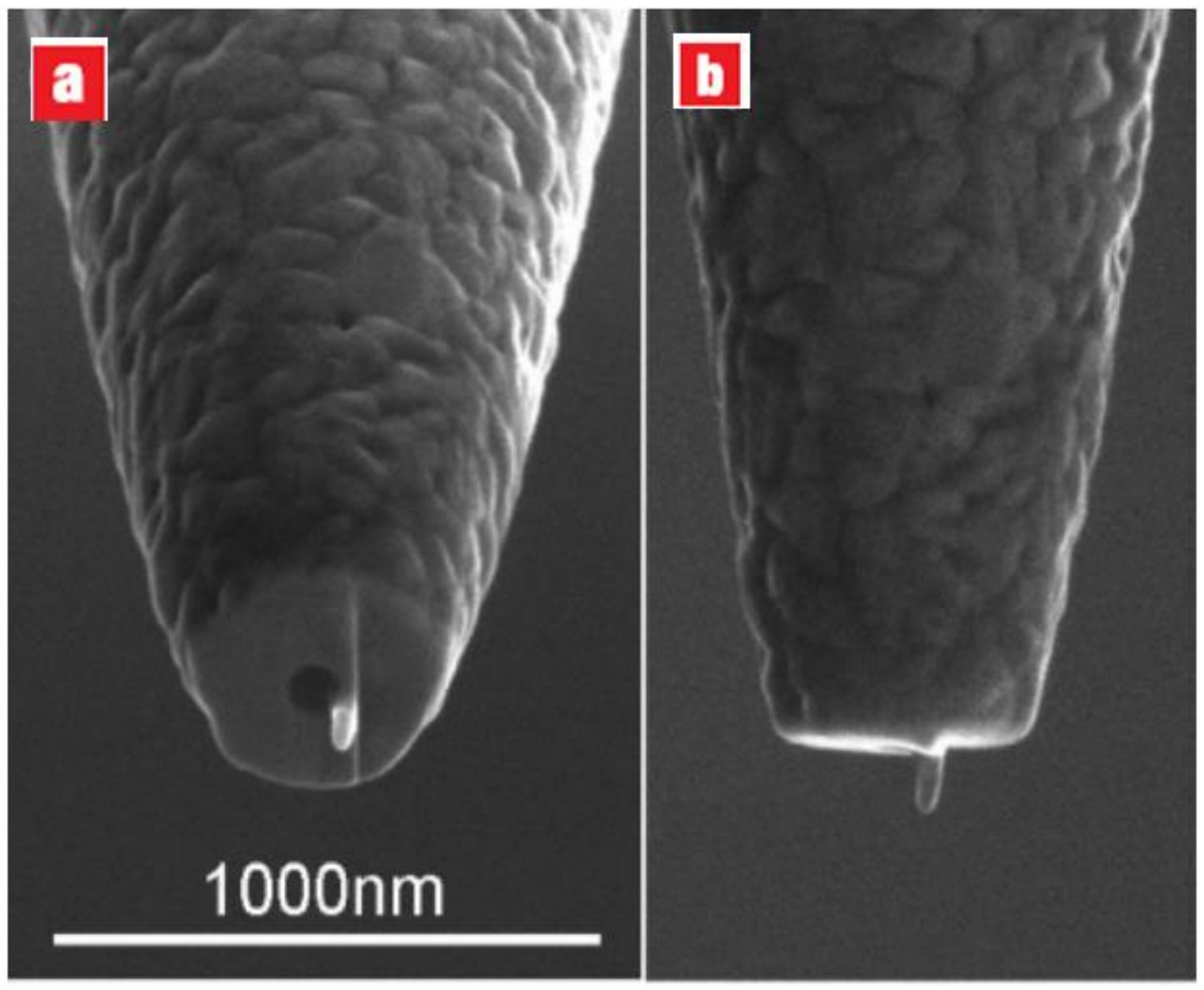

Fig. 16. A probe-based nanoantenna (SEM images): (a) viewed from a 52° angle and (b) side view [52].

For plasmonic microsystems applications, smooth surfaces are of utmost importance. Unfortunately, metal films deposited by evaporation are inherently rough due to polycrystallinity. In order to overcome this drawback, Prof. Oh's research group from the University of Minnesota combined accurately patterned silicon substrates with template stripping to achieve ultra-smooth metal films with grooves [53]. Fig. 17 presents the process in which a silicon substrate with circularly patterned concentric grooves are defined by FIB. A 275-nm thin film Ag was thermally evaporated on the silicon substrate, next, epoxy was added, and peeled off the bilayer.

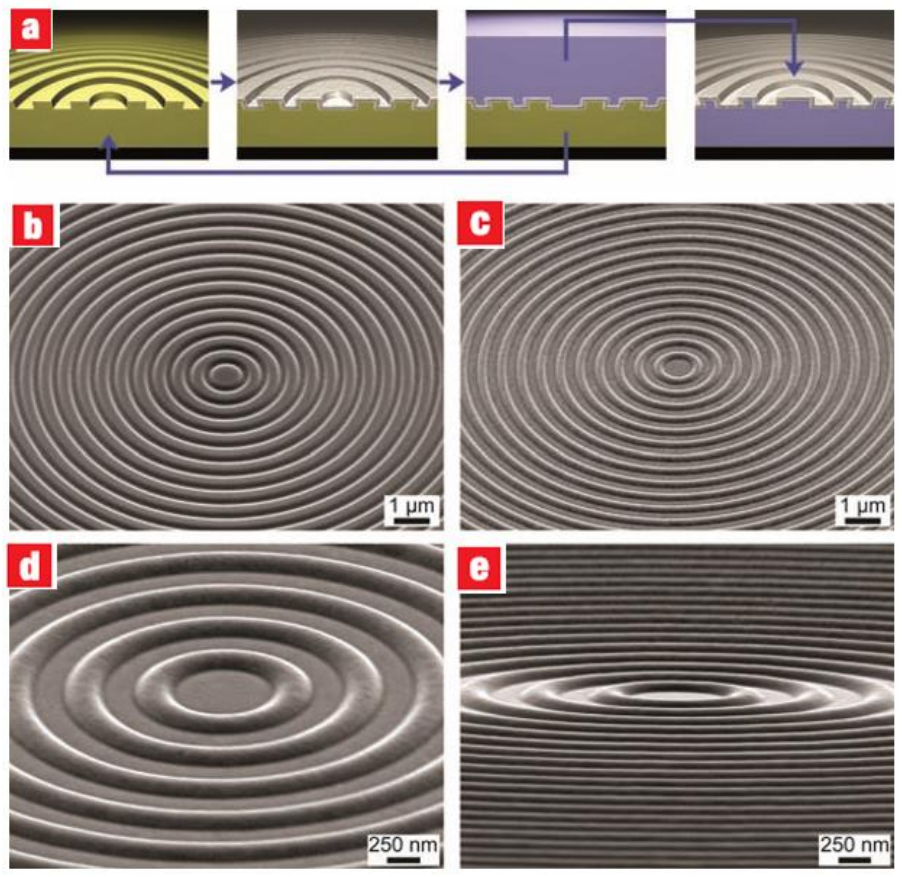

Fig. 17. Ultrasmooth plasmonic patterning: (a) Schematic of the fabrication process. (b) FIB patterned silicon wafer, (c to e) SEM images of the ultrasmooth patterns after separating the silicon template from the circular grooves [53].

### *C. Scanning Probe Lithography (SPL)*

Up to this point, we have reviewed sub-100-nm and photonic/plasmonic nanofabrication methods based on photons and charged beams (electrons and ions). To be able to fabricate sub-100-nm scale plasmonic microsystem structures using photolithography, many tricks apart from short wavelength have to be used in photon-based lithography (such as high numerical aperture (NA)). In addition, very complicated charged beam systems are required in charged particle-based lithography. Another problem with lithography using either photons or charged beams is that they always rely on a polymer material (photo-resist or electron-resist) as an imaging layer.

For low-cost and ultrahigh resolution nanoscale patterning and printing technologies, scanning probe lithography (SPL) is an alternative to expensive photon or charged beam techniques because it does not require costly equipment. SPL utilizes a scanning probe device (a sharp tip) in close vicinity of a sample to draw nanometer-scale features on the sample. Scanning probe microscopy (SPM) refers to a group of probe-based methods and techniques that examine the local interaction between a sharp tip (less than 100 nm in radius) and a sample to attain electrical, mechanical, biological, or chemical information about the surface with high spatial resolution. Today there are many different types of SPMs used for diverse applications ranging from biological probing to material science to semiconductor metrology [54]. Three main technologies inside the SPM family are scanning tunneling microscopy (STM) [55], atomic force microscopy (AFM) [56], and NSOM [57]. Among them, NSOM is the most common tool used for plasmonic fabrication. In NSOM, patterning of nanostructures is done via a direct writing process that works by the confined optical near-field at the tip of a fiber probe. The fabrication of nanostructures in a size below the diffraction limit of the light source is attainable by using localized plasmonic waves. Both contact mode [58] and non-contact mode [59] NSOM has been used in order to fabricate sub-100-nm elements for plasmonic and 2D photonic structures in [60].

The first practical plasmonic NSOM probe was first demonstrated in 2008 by the pioneering work of Wang et al. [61]. The conic plasmonic lens they demonstrated consists of a subwavelength aperture at the apex of the cone enclosed by concentric rings in an aluminum thin film deposited on a tapered fiber tip, as shown in Fig. 18a. By using this method, the tight focus of an approximately 100-nm beam spot was obtained.

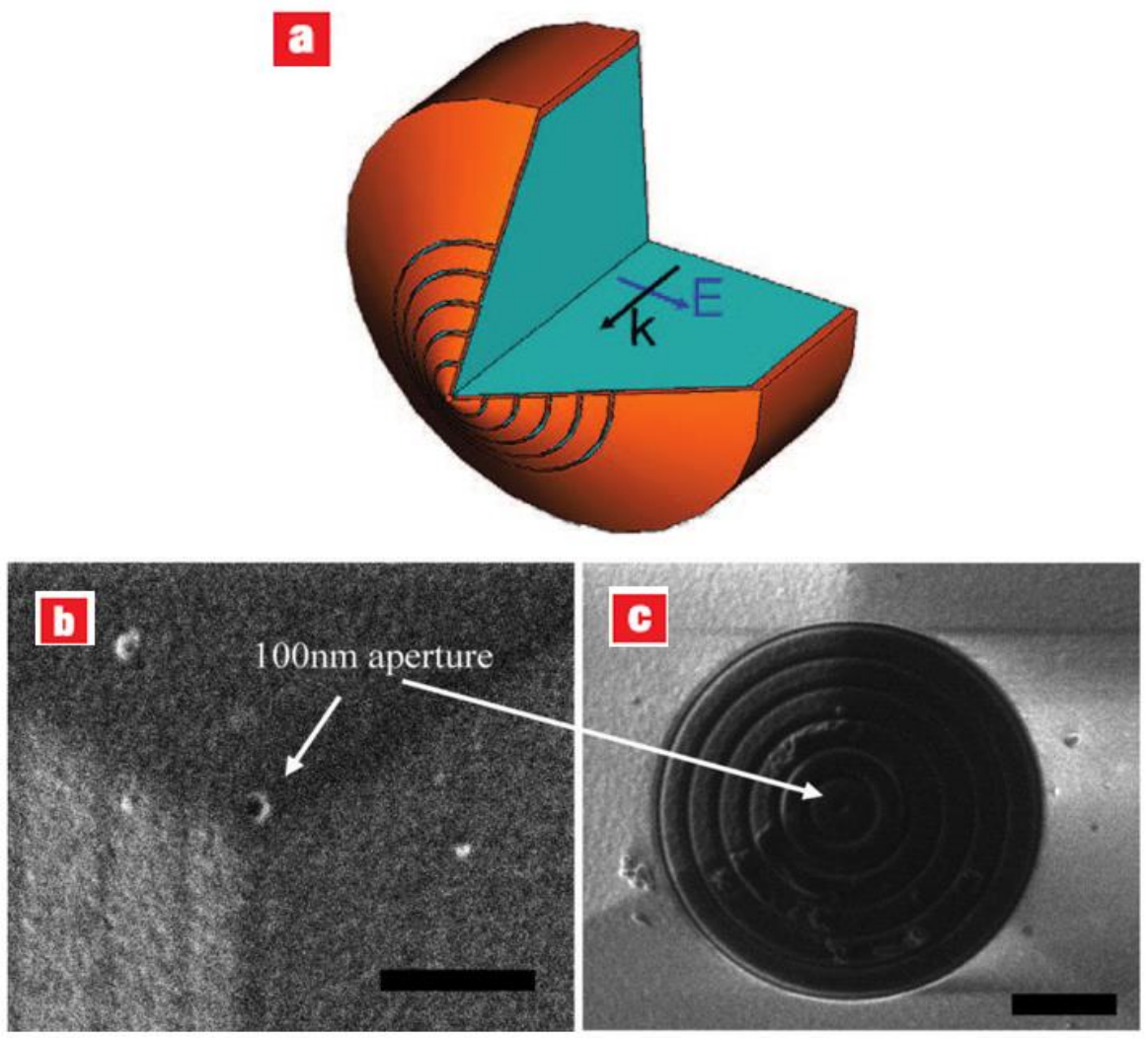

Fig. 18. (a) Schematic drawing of the aperture based conic plasmonic lens. NSOM probe consists of nanostructured plasmonic lens being fabricated on the end of an optical fiber. 100 nm aperture on the NSOM tip (b) before and (c) after the fabrication of the plasmonic lens. Scale bar is 1 um [61].

One of the most attractive and promising techniques as far as plasmonics is concerned is the dip-pen nanolithography (DPN), which first was reported in 1999 [62]. DPN employs an AFM tip which is coated with a thin film in order to transfer molecules from the tip to a solid substrate of interest (such as gold) through capillary transport. The mechanism can be compared to writing with a fountain pen on paper in which the ink is transported from the pen to a substrate. The advantage of DPN over other nanolithography techniques is the potential to selectively place several types of molecules to the same specific site of a nanostructure, providing the capability to modify selective chemical functionality of the surface in nanoscale. Accurate control of DPN parameters including, temperature, humidity, writing speed, and tip substrate, can be used to obtain dots and lines as small as 10-15 nm [63].

### *D. Emerging techniques*

Although costly nanofabrication tools such as EBL, FIB, or EUV projection lithography can be successfully implemented to construct a variety of plasmonic microsystems structures, they do not offer expedient, low cost fabrication methods that allow for industrial-scale manufacturing. Alternative techniques to cost-intensive or limited-access fabrication methods with nanometer resolution have been under development for nearly three decades. This section will briefly review the most recent and state of the art sub-100 nm plasmonic microsystem structures nanofabrication techniques.

#### *1) Nanoimprint lithography (NIL)*

Nanoimprint lithography (NIL) is one of the most promising low-cost, high-throughput technologies for plasmonic microsystems fabrication [64, 65]. Its main element is a patterned "mold" or "stamp" that is pressed onto the surface of a polymer in order to pass on its pattern. In 1995, NIL was proposed and demonstrated as a technology for sub-50 nm nanopatterning [66]. There have been several key achievements in the development of NIL as a nanofabrication technique over the past several years. Sub-100 nm resolution metallic nanograting patterns on a 6-in. silicon substrate was reported in 2000 [67]. In the same year, a new class of specific polymers developed for NIL which triggered its commercial use [68].

NIL can be used to develop various plasmonic geometries including grooves, slits, holes, and 3-dimensional tiered structures. Recently, the ability to fabricate fishnets by nanoimprinting has been demonstrated [69]. The structure was fabricated in a pre-deposited three-layer MDM stack. J. L. Skinner et al. in [70] explored the possibility of plasmonic biosensing using silver and aluminum nanohole arrays. Both arrays, with periodicity of 500 nm and nanohole diameters of 110 nm, were fabricated through NIL. Fig. 19 shows arrays of gold nanocones that have been fabricated using UV-NIL for plasmonic sensing applications [71]. Nanocones are 130 nm in base diameter and are ordered in a square grid with a 300-nm period.

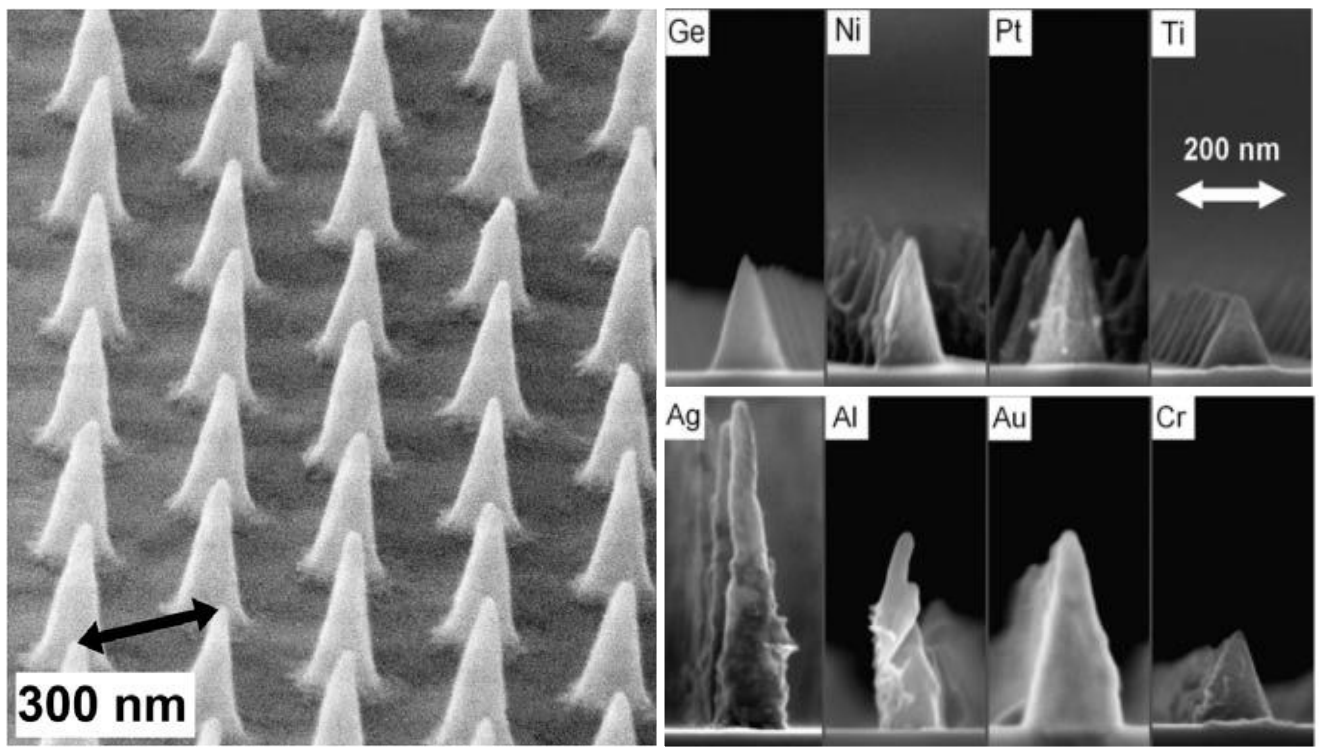

Fig. 19. An array of gold nanocones on a silicon substrate. The period is 300 nm, the cone bottom diameter is 130 nm, and the average height is 257 nm (Ti/Au 20/230 nm). The proposed method can be used to fabricate a variety of plasmonic metallic structures [71].

For plasmonic microsystem applications, sharp V-groove structures are of a great importance since they guide and propagate "channel plasmon-polariton modes". Kristensen et al. [72, 73] have exploited a combination of UV-NIL for the fabrication of sharp V-groove structures. The fabricated structures are presented in Fig. 20.

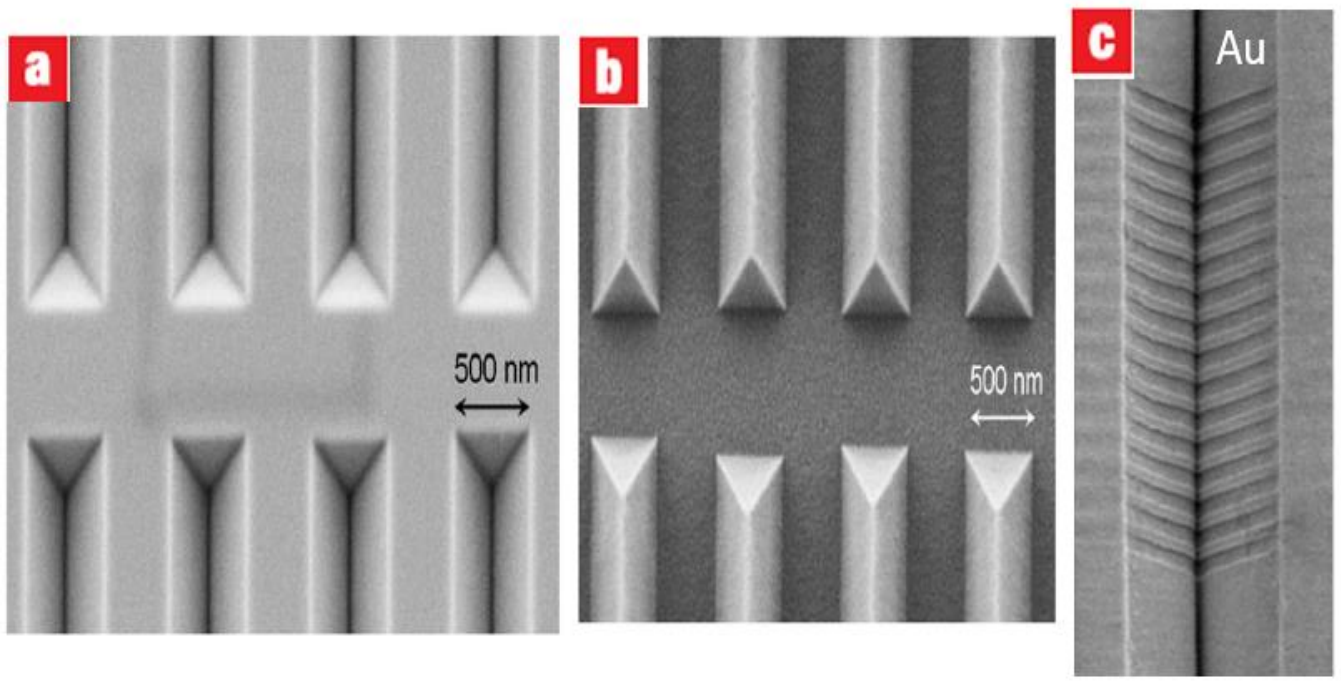

Fig. 20. (a-b) Silicon stamp, and its replication in PMMA. (c) Tilted SEM image of a V-groove containing a Bragg grating filter (BGF) [72].

The authors later used the same methodology to design a spectral plasmonic filter with Bragg gratings in V-groove waveguides [72].

*2) Soft lithography*

In the NIL section, we stated that molded polymer pattern structures can be transferred from a mold by pressing it into a substrate by NIL. In 1993, Kumar and Whitesides proposed another imprinting technique which employs a flexible elastomer soft mold (polydimethylsiloxane (PDMS)) to transfer alkanethiol ink to a substrate coated with gold thin film [74]. The proposed method was later called soft lithography and became an important replication technique in plasmonic microsystems. The introduction of novel composite stamps in 2002 extended the capabilities of soft lithography to the generation of 50-100 nm features. Whiteside's research group in 2002 developed an improved soft mold, namely "composite stamp"; a two-layer stamp devised of a stiff layer (30-40 µm *h*-PDMS) supported by a thick flexible layer (3 mm slab of 184 PDMS) [75] which extended the capability of soft lithography to sub-50 nm features. Fig. 21 shows lines with about 50 nm thickness fabricated using the proposed composite stamp.

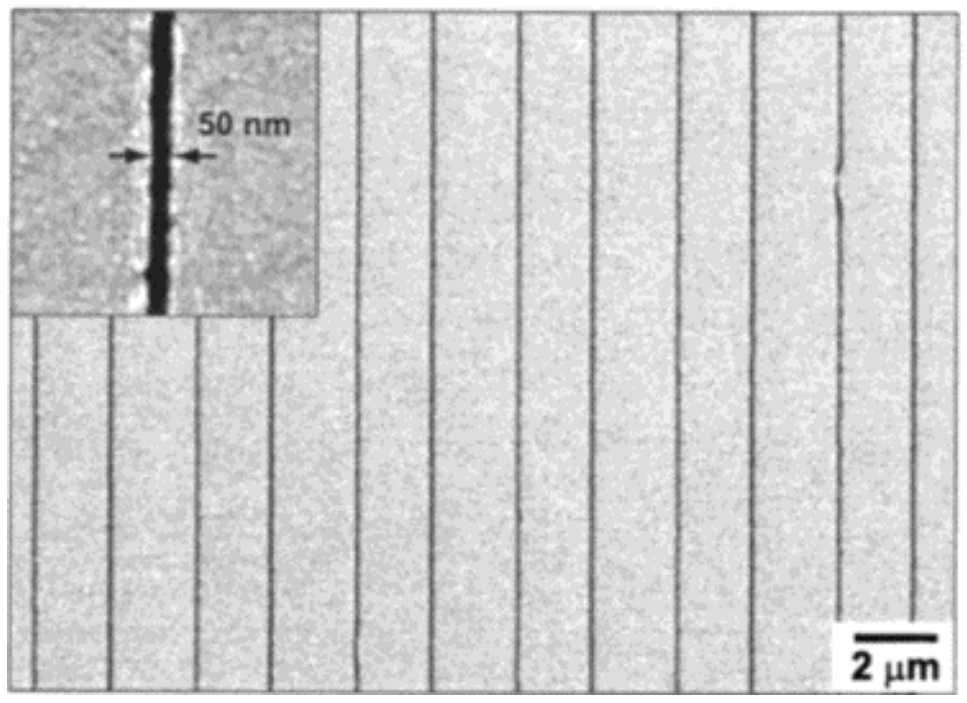

Fig. 21. SEM of the fabricated uniform, 50 nm thick lines manufactured from the proposed composite soft stamp. Figure taken from [75].

While soft lithography can pattern plasmonic metals over large areas, this technique has some drawbacks in producing ultra-smooth metal patterned that are required for plasmonic microsystems. Besides, periodic gratings with less than approximately 80 nm thickness, adhere to each other because of the displacement during stamp alignment and they cannot reform into the correct order because of adhesion forces.

In 2007, Teri Odom's group developed soft interference lithography (SIL): a high-throughput nanofabrication technique based on soft lithography applied to plasmonic microsystem applications [76]. SIL merges the wafer-scale nanopattern production capabilities of interference lithography (IL) [77, 78] with the flexibility and versatility of soft lithography.

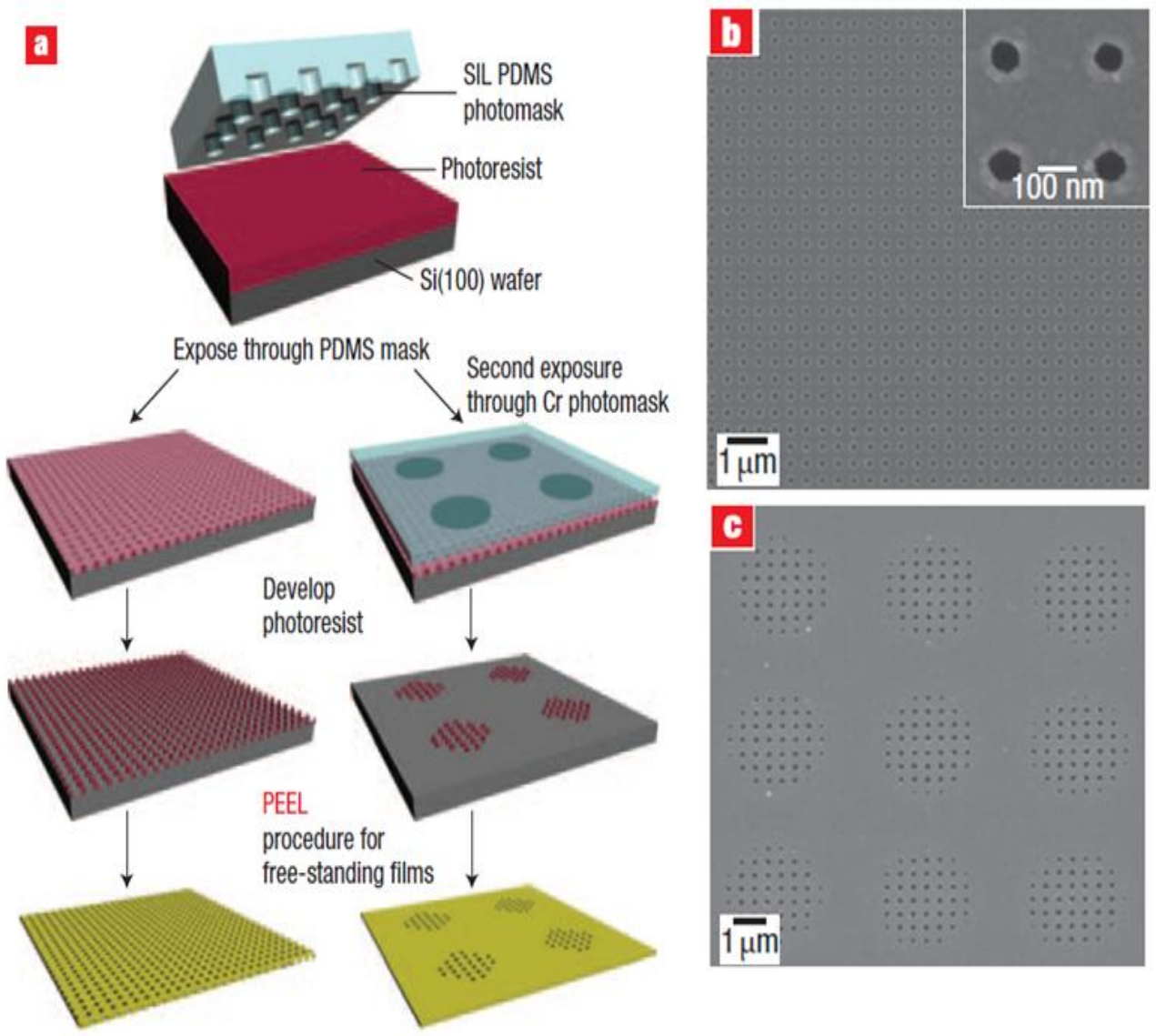

Fig. 22. (a) Illustrating the SIL fabrication process of infinite nanohole arrays and finite-sized patches of holes. (b) Infinite silicon nanohole arrays with 100-nm size holes. (c) Silicon patch nanohole array [76].

SIL uses nanoscale patterns generated by IL as high-resolution stamps for soft lithography, as shown in Fig. 22. To transfer patterns from the master onto a positive-tone photoresist, the SIL photomask is placed in contact with the photoresist and then exposed to UV light. Developing the photoresist at this point generates posts with identical tangential dimensions to the IL master (that is, infinite arrays). Exposing the photoresist for a second time through a chromium mask and subsequently developing the pattern forms microscale patches of post arrays (referred to as finite arrays). The photoresist patterns are mapped onto metal or dielectric films using a soft nanofabrication procedure called PEEL (Phase-shifting photolithography, Etching, Electron-beam deposition, and Lift-off) [76]. Fig. 23 displays nanopyramidal gratings for screening plasmonic materials [79] fabricated via SIL. The structure has been used to generate plasmon dispersion diagram for Al, Ag, Au, Cu, and Pd.

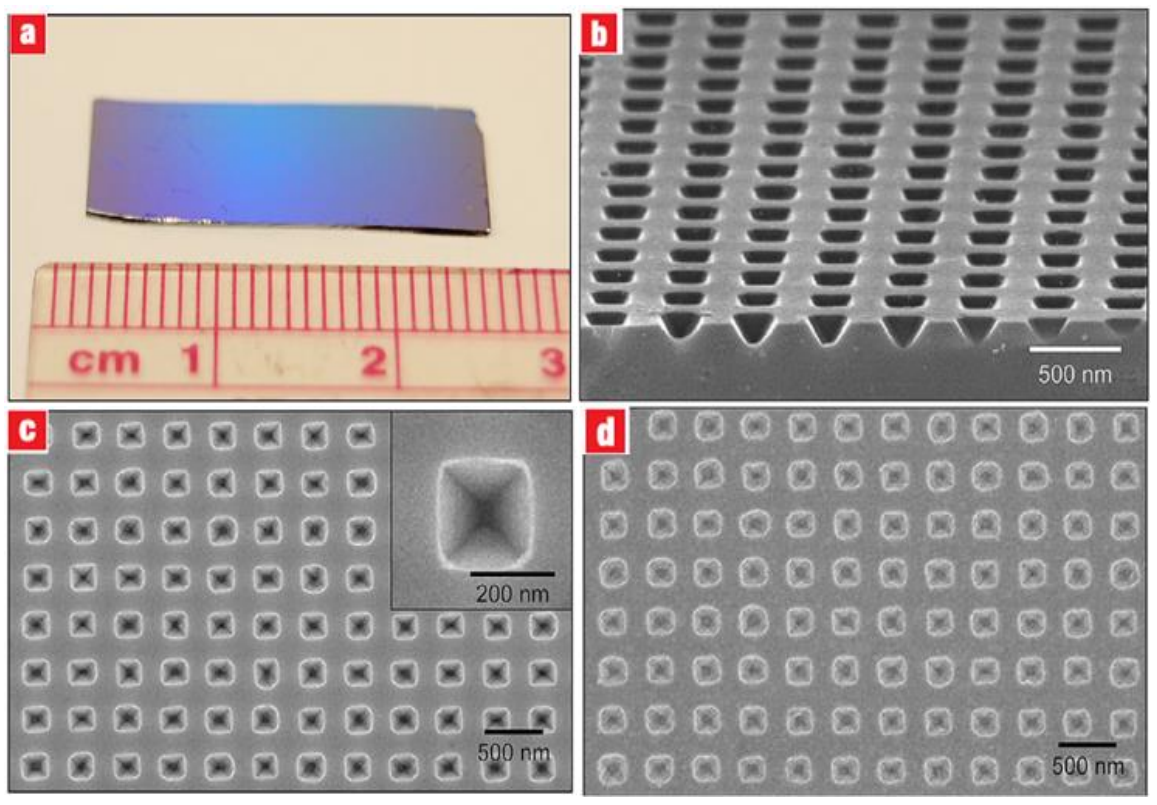

Fig. 23. Plasmonic materials on pyramidal nanograting. Optical micrograph (a) SEM (b and c) images of a large-area (0.9cm × 2.3cm) silicon pyramidal slit-grating. (d) SEM image of a 170-nm-thick Au film deposited on the silicon master. Reprinted from [79].

### 3) *Nanosphere lithography (NSL)*

To enhance the consistency of particle size and arrangement, in 1995 Van Duyne and colleagues proposed a unique and simple fabrication method to fabricate plasmonic metal nanoparticles called nanosphere lithography (NSL) [80]. NSL utilizes a monolayer of tightly packed polystyrene spheres on a substrate surface as a masking layer which can be patterned using both bottom-up and top-down techniques such as EBL or self-assembly [81]. The first step in NSL is dropping polystyrene nanospheres on a pristine, pre-prepared glass substrate. The hexagonally close-packed (Fischer pattern) nanospheres create a crystal structure in which the gaps between the spheres form a regular array of dots. Next, the array is filled in with thermally evaporated silver.

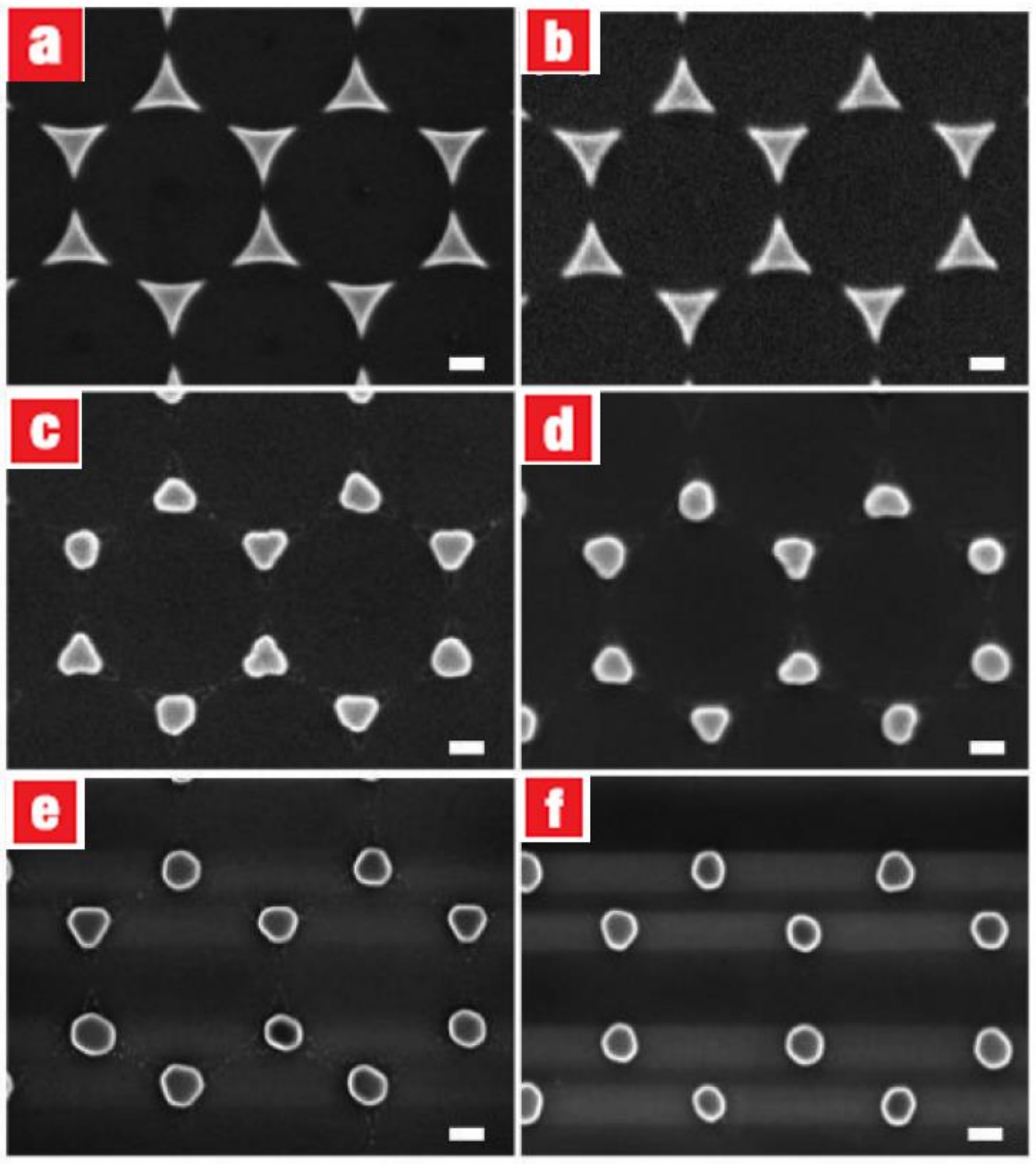

Fig. 24. SEM images of (a) deposited and (b-f) annealed Au nanoparticles. The pattern was annealed gradually with increasing temperatures at each phase, the images were taken after each annealing step. The corresponding annealing temperatures were (b) 205, (c) 325, (d) 451, (e) 700, and (f) 930 °C. Scale bars = 200 nm [82].

After deposition, the polystyrene spheres are removed by agitating (sonicating) the entire substrate in either $CH_2Cl_2$ acid or absolute ethanol, and the product is an array of triangular dots. Fig. 24 shows the triangular nanoparticle shape after deposition by NSL [82]. As it can be seen from the results, different shape nanoparticles with different sizes has been fabricated using the proposed method. The method can be readily extended to pattern plasmonic microsystem devices [83, 84].

NSL has been widely used for the fabrication of plasmonic patterns such as nanowires, nanocones and triangular arrays [85-87]. As an interesting example, Fig. 25 shows an array of gold nanocones fabricated with NSL over an area larger than 100 $\mu m^2$.

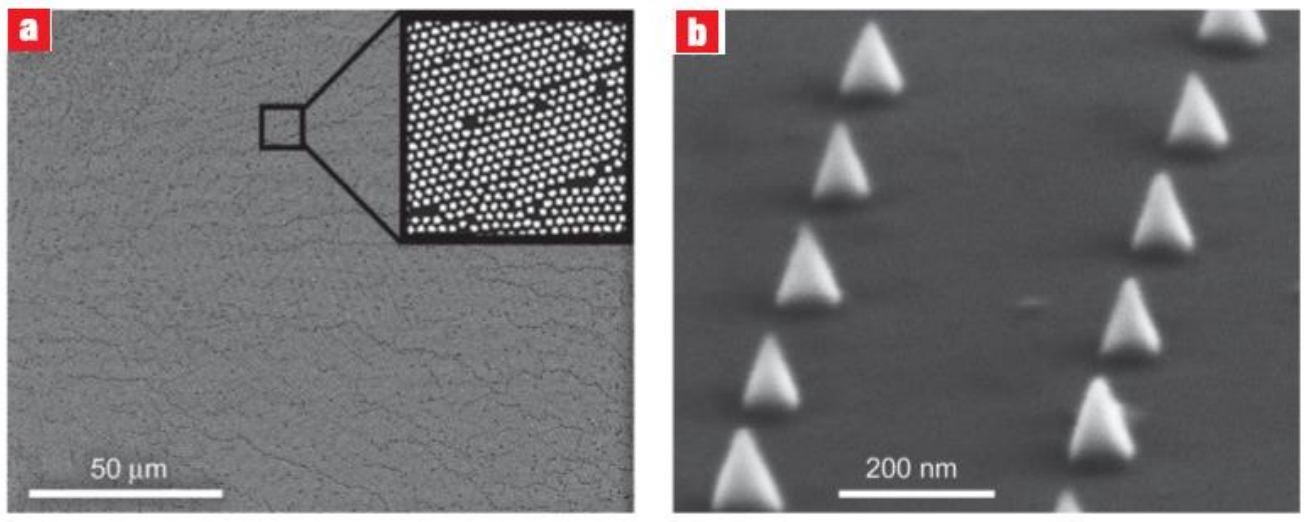

Fig. 25. (a) SEM outline of gold nanocones fabricated over an area of more than 100 µm$^2$. Polystyrene spheres with diameters decreased in an oxygen plasma were exploited as the mask. The magnified image presents the homogeneous hexagonal pattern of cones with individual grain boundaries and missing cones (b) Procedure (double layer and alumina mask) utilized to a silver film leads to sharp-tipped silver cones with a smaller tip aperture angle than for the gold cones [88].

In this example authors reported the homogeneous distribution of gold nanocones with the tip radius of less than 10nm using a monolayer mask of oxygen plasma treated polystyrene nanospheres.

## IV. Applications

Based on the theories discussed in section I and the versatile fabrication techniques described in section II, plasmonic microsystem devices have been utilized in different applications such as sensing, optical data storage, microimaging and displaying. In the following section, we will discuss the use of plasmonic microsystems in the following applications: biosensing, mainly on the refractive-index based label-free biosensing; plasmonic integrated lab-on-chip for Point-of-Care (POC) systems; plasmonic for near-field scanning optical microscopy, and plasmonics on-chip systems for cellular imaging.

### *A. Nanostructured plasmonic label-free biosensors*

The unique optical properties of plasmon resonant nanostructures make them ideal for label-free biosensing applications [89-92]. The basis of the plasmonic-based sensing mechanism is that the surface plasmonic resonance (SPR) is highly sensitive to the refractive index of the medium, and the biomolecules attached to the biosensor lead to a change in the dielectric environment (i.e. refractive index) at the surface, which eventually leads to a change in SPR resolved in angle, wavelength or light intensity. This mechanism makes plasmonic sensors a label-free biosensor, meaning that once the analyte is captured on the sensor surface, no additional tags are required for the detection of the binding events. Also, plasmonic sensors are robust for real-time monitoring of biomolecules interactions. Biosensors based on propagating SPR in metallic films are the most well-known and commercially available optical biosensors. Film-based SPR sensors have become the “gold standard” for characterizing biomolecule interactions. Excitation of the SPR in thin films is usually carried out in the so-called Kretschmann configuration, in which light is coupled into the gold film by a prism that facilitates total internal reflection. Fig. 26 presents the most common SPR sensing experiment process [91].

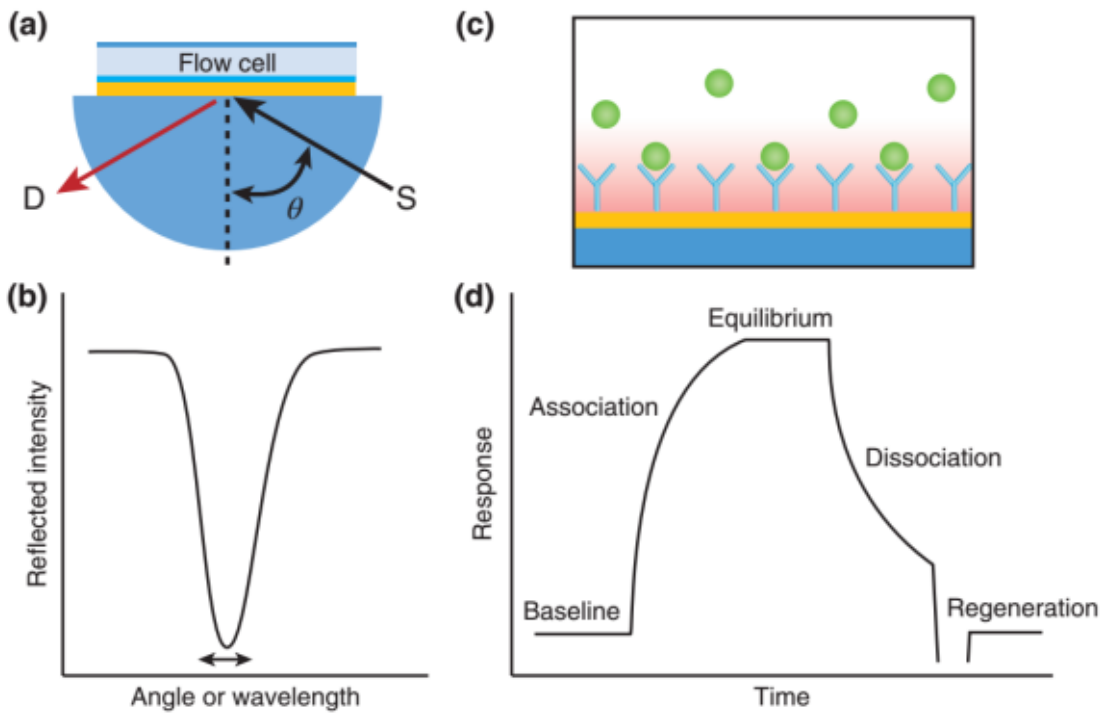

Fig. 26. Film-based SPR sensor [91]. (a) The most commonly used Kretschmann configuration, S: Source, D: Detector. (b) The detection presenting as a change in SPR angle or wavelength. (c) A schematic showing target analyte binding by the receptor immobilized on the film surface. (d) A typical SPR monitoring curve and the kinetic parameters that can be calculated from the curve.

First, a bioreceptor (antibody, aptamer, DNA, etc) is immobilized to the surface using either physisorption or covalent binding. Subsequently, a baseline response of the instrument is established. After a target analyte is introduced to the surface, change in SPR angle or wavelength can be monitored over time. The SPR response can be used to determine kinetic parameters of the binding interactions. However, film-based SPR sensors have several disadvantages. For example, a light coupling mechanism is required so that the momentum of incoming photons can match to the free surface electrons of the gold film and create a SPR. Also, a temperature and vibrations have to be controlled precisely to produce a stable SPR signal. These requirements make the SPR instrument bulky and difficult to incorporate into a compact and miniaturized sensing system that can be applied in a lab-on-chip setting. Plasmonic sensors based on localized SPR (LSPRs) have the potential to be used as small, portable and multiplexed sensing devices compared with film-based SPR sensors. LSPRs are based on metallic nanostructures that are significantly smaller than the wavelength of incident light, so that SPRs are only around the nanoscale objects and nanoparticles. The spectral signatures of LSPRs can be tuned by varying the shape, size and composition of the metallic nanostructures. This feature is advantageous for biological sensing, because the detection wavelengths can be tuned to avoid overlapping with the spectral features of some natural biological chromophores, such as hemoglobin in blood samples, so that the sensitivity can be improved [93].

In a notable example of biosensing for diagnostic application, an LSPR sensor has been employed to detect a biomarker for Alzheimer’s disease, amyloid-beta-derived diffusible ligand (ADDL), in both synthetic and human patient samples [94] (Fig. 27).

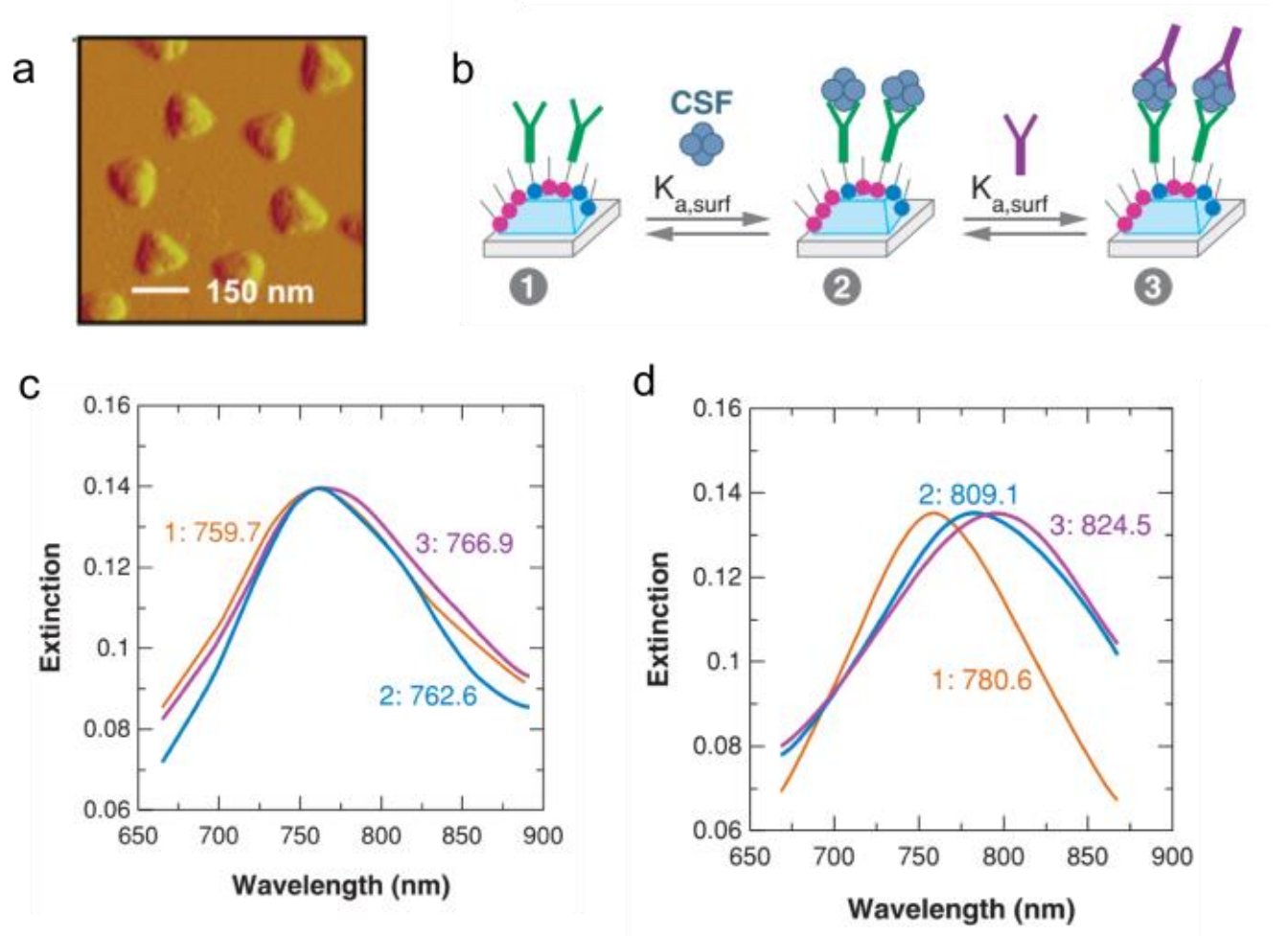

Fig. 27. LSPR sensor for the detection of a biomarker for Alzheimer's disease [90]. (a) AFM microscopy shows the image of the fabricated Ag nanoparticles on mica substrate using NSL [94]. (b) Surface chemistry for ADDL detection in human cerebrospinal fluid (CSF) using antibody sandwich assay. The steps include (1) surface immobilization of anti-ADDL, (2) the incubation of human CSF, and (3) the introduction of the second capping antibody. (c) LSPR spectra for each step in (b) for an age-matched control patient. (d) LSPR spectra for each step for an Alzheimer's patient.

In this work, triangular Ag nanotriangle arrays were fabricated on mica substrate by NSL. For the first time, LSPR technology was used to monitor a sandwich assay. First, the nanoparticle array was functionalized with antibodies specific for ADDLs; next, the surface was exposed to ADDLs for the binding event to occur; finally, a second capping antibody was introduced to bind to the surface and complete the sandwich assay. UV-vis spectroscopy was employed to monitor the extinction signal after each modification step and the binding event. Initial experiments showed that synthetic ADDL concentrations on the order of 100 fM could be detected. Fig. 27c-d shows the detection using CSF from both an Alzheimer's patient and an age-matched control. It shows that the LSPR shifts were significantly larger for the patient sample after the CSF exposure and the following adding of the second capping antibody. The sensitivity can be further enhanced if the molecular resonance of the analyte or the second antibody overlaps with the LSPR of the nanoparticles. By using an associated biomarker and antibody pair, plasmonic biosensing will be a promising technology for the diagnosis of any diseases.

Recently, nanohole array-based plasmonic sensors have been shown as one of the most exciting platform for label-free, high-throughput and miniaturized sensing devices [95, 96]. Nanohole array plasmonic sensors is based on the effect of extraordinary optical transmission (EOT). EOT stems from SPR that is excited by the grating orders of the pore arrays, so that the amount of transmitted light at certain wavelengths is much larger than predicted by classical aperture theory. The spectra are sensitive to the size and shape of the pores and the dimensions of the array. Any changes in the local refractive index caused by target binding events will lead to changes in optical signatures. Nanohole arrays can work as a sensor either by monitoring the light intensity changes or the spectrum shift passing through the pores.

There are several advantages of nanohole arrays that make them so promising for biosensing. First, the sensor can be measured using a simple collinear optical configuration, which makes the instrumentation simple and portable. Second, the sensor can be integrated into a small chip, so it can be used for miniatured and multiplexing detection system. In addition, it shows that by flowing analytes through system, the limit of detection can be improved significantly, as the flow-through configuration solves the mass-transport problem on the sensor surface [97, 98]. This is especially important for detection of extremely low concentration analytes. Fig. 28 shows a handheld nanohole array sensing device which has been used for high-throughput and imaging based sensing [99]. Fig. 28a is SEM images of the nanohole arrays in 6 sensor pixels of size 100 × 100 μm. The diameter of the nanohole is 200 nm. Fig. 28b is a photo of the portable device, weighing 60 g and 7.5 cm tall, which was designed for Point-of-Care (POC) applications. Fig. 28c shows the schematic of the on-chip imaging platform, which only contains a battery, an LED, a plasmonic chip and a CMOS imager chip. For biomedical applications, nanohole array plasmonic sensors have been recently applied for the detection and molecular profiling of tumor-derived exosomes, which is a potential circulating biomarker for "liquid biopsy" cancer diagnostics [100]. Nanohole array sensors were used to analyze exosomes from ascites samples from ovarian cancer patients, which can be differentiated from healthy controls.

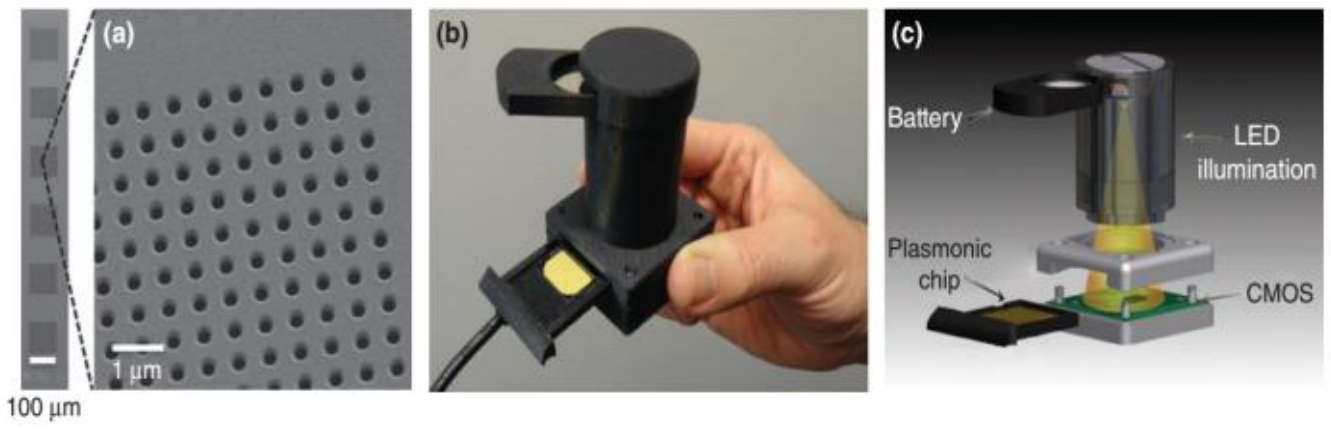

Fig. 28. Nanohole array plasmonic sensor [99]. (a) SEM images of the nanohole arrays in 6 sensor pixels of size 100 × 100 μm. The diameter of the nanohole is 200 nm. (b) A photo of the portable device, weighing 60 g and 7.5 cm tall, designed for POC applications. (a) Schematic of the on-chip imaging biosensing platform contains a battery, an LED, a plasmonic chip and a CMOS imager chip.

### *B. Microfluidics-enabled plasmonic chips for biosensing and imaging*

Microfluidics have been widely used in sample isolation, preparation, analysis and delivery [101-103]. The integration of plasmonic sensing and microfluidic technologies provides the opportunity to build label-free biosensors on a lab-on-chip (LOC) platform, which can be potentially used in the global health, primary care and POC applications. Combining microfluidics with plasmonics combines advantages from both of these fields. On one hand, light can be used to direct the motion of fluids in microfluidic chips. For instance, optical tweezers have been utilized to build valves and pumps to induce flow in microfluidic channels [104, 105]. On the other hand, fluids can be used to change optical parameters in plasmonics [106]. Current state-of-the-art plasmonic-based biosensors have been developed for the applications using the platforms such as SPR, LSPR and SPR imaging.

An example of an LSPR-based biosensor with fully integrated microfluidics is shown in Fig. 29 [107]. A periodic arrangement of gold nanorods was immobilized on a glass substrate by nanofabrication procedure. The sensor was then integrated with a PDMS-based microfluidic device controlled by micromechanical valves. The system provides parallel, real-time monitoring of 32 sensing sites across 8 independent microfluidics channels, which later enables the screening of a variety of biomarkers. The optical setup used to monitor the chip consisted of a homemade microscope in a bright-field transmission configuration, which was equipped with a scanning detection system, a visible−near-infrared light source and a spectrometer (Fig. 29c). The authors demonstrated the platform for cancer diagnostics through fast detection of cancer biomarkers human alpha-feto-protein and prostate specific antigen, down to concentrations of 500 pg/ml in a complex matrix consisting of 50% human serum.

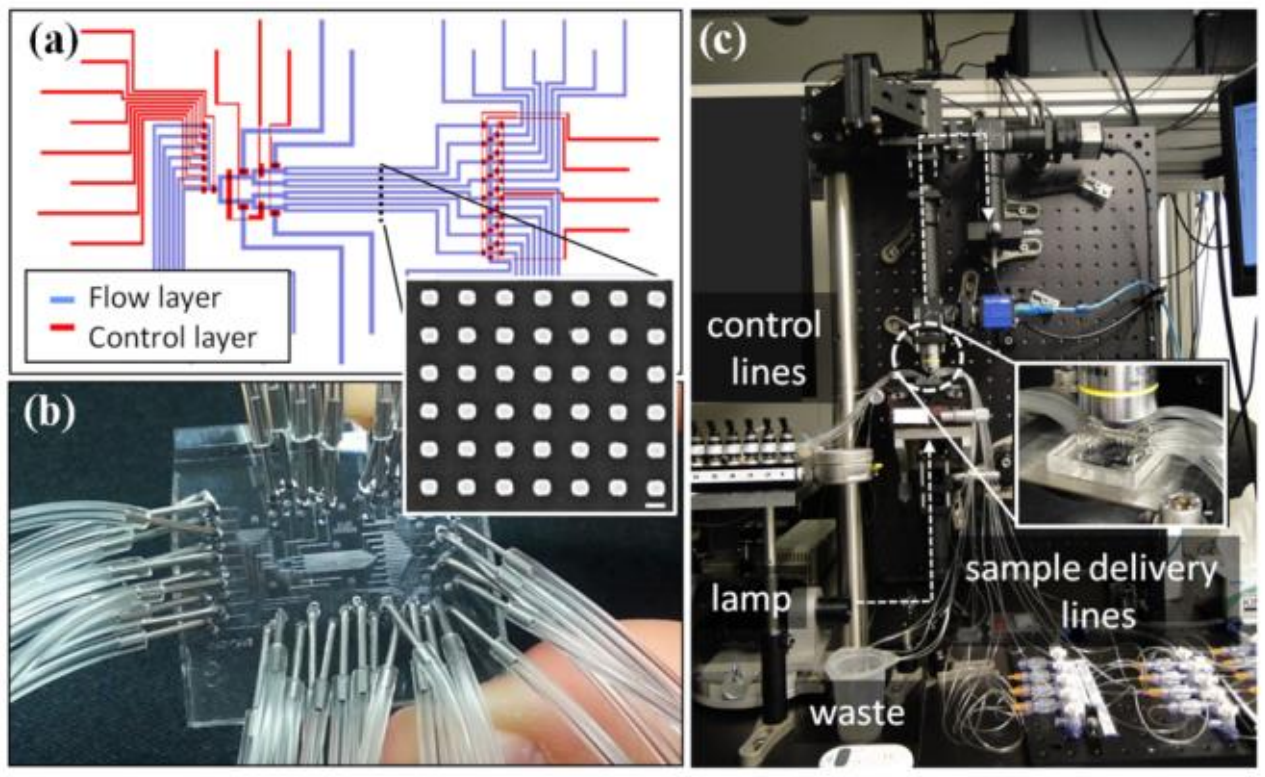

Fig. 29. LSPR-based biosensor integrated with microfluidics [107]. (a) Schematic of the flow and control layers of the microfluidic chip. (b) A photo of the actual chip connected to tubes. The inset shows a SEM image of the plasmonic gold nanorod array. Scale bar 200 nm. (c) A photo of the optical setup for readout.

SPR imaging is a high-throughput method which allows imaging-based detection. Based on SPR techniques, the binding event of analytes on the sensor leads to local refractive index changes, and this lead to a change in the spatial light intensity. Therefore, the captured images allow for both temporal and spatial monitoring of surface binding events in a label-free format. Microfluidic chip has been integrated with SPR imaging. For example, microfluidic chips with channels, valves, pumps, flow sensors and temperature controller was developed using MEMS technology. The system has been demonstrated in the detection of interaction between IgG and anti-IgG [108].

Recently a microfluidic chip has been developed for immunoassay-based SPR imaging [109]. Fig. 30a shows a schematic of the microfluidic device. The system contained two layers. The flow layer was comprised of a crossed-flow structure, which allows for the loading of two reagents at the same time and the occurrence of the immunoreaction. Another control layer contains microfluidic pumps for the control of liquid flow. The intersection of the flow layer was located on the top of an array of gold spots, which were used as SPR sensing substrate for multiplexed detection of immunoreactions. To demonstrate the performance of the chip, the binding of anti-biotin antibodies to surface-immobilized biotinylated bovine serum albumin was monitored, with and without a signal amplification step. Real-time immunoassay imaging allowed for monitoring of immunoreaction on each gold spot in ~10 min, and the system achieved a sub-nanomolar detection limit. Subsequently, a two-step immunoassay was performed for signal amplification, by delivering another anti-goat IgG antibody labelled with gold nanoparticle to the gold spots. This process improved the limit of detection down to ~40 pM, but the time increased to ~60 min. With the small dimensions of 2-3 cm and less than 100 pM detection performance, this SPR imaging system has the potential to be used in POC applications.

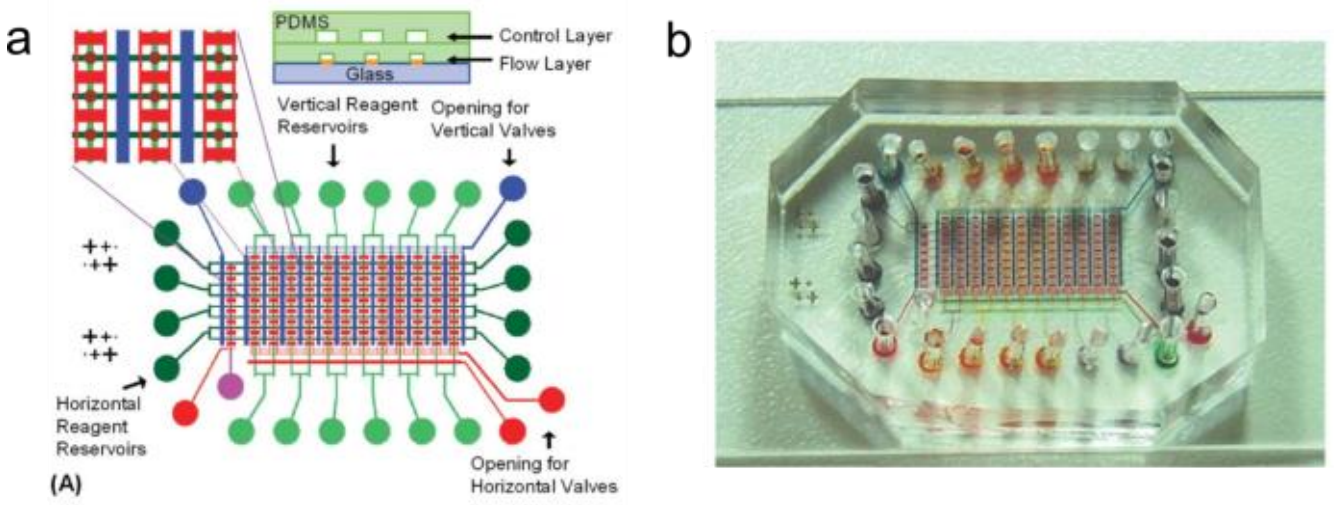

Fig. 30. A microfluidic SPRi platform [109]. (a) The schematic of the chip, where the immunoreaction takes place in the lower layer, and the upper layer controls the fluid flow by microfluidic pumps. (b) A photo of the microfluidic device with dye fluidics.

### *C. Plasmonics for near-field scanning optical microscopy and sensing*

Optical microscopy is an important technique which can be used for imaging-based biosensing. An optical microscope has a lot of advantages, and becomes much more convenient if it can be used to see a sample in nanometer resolution. However, this goal is limited because of the diffraction of light, and the spatial resolution of a conventional optical microscope is about 0.5 μm. However, plasmonic nanostructures make it possible to examine system with spatial resolution beyond the diffraction limit of light, using a nano-sized metallic probe tip to scan the sample surface. The mechanism is based on the excitation of localized modes of surface plasmon polaritons (SPP) at the metallic tip, which generates a nano-sized light spot as the light source of an optical microscope. Therefore, extremely high spatial resolution can be achieved. According to Abbe's diffraction theory, the major roadblock in conventional optical systems, the wavelength of light ($\lambda$) has to be reduced to obtain a higher resolution. The wavelength of light is inversely proportional to the angular frequency, $\omega$, as $\lambda=2\pi c/\omega$, in which $c$ is the speed of light. Therefore, the speed of light can be reduced to shorten the wavelength of light. In conventional optical devices, this has been done by filling the space between the lens and sample with high-refractive-index materials such as immersion oil. However, when increasing the refractive index further, it is difficult to keep the medium transparent. SPPs can focus the optical energy of light independent of the wavelength. SPPs are a form of evanescent wave on the metal surface associated with the collective oscillation of free electrons, which can be used for high-resolution imaging.

NSOM is a technique that uses SPPs on a metal probe with dimensions much lower than the wavelength. In NSOM, resolution is defined by the aperture/tip size rather than the incident wavelength. NSOM converts the non-propagative near-field signal into a measurable far-field contribution, so that nanoscale resolution can be achieved with optical imaging. NSOM techniques can be classified into the "aperture" and "apertureless" categories. Aperture NSOM uses an optical fiber with a metalized nanoaperture; while apertureless NSOM uses a metallic tip as a nanoantenna to provide higher spatial resolution. Fig. 31 shows the schematics of the two NSOM techniques [110].

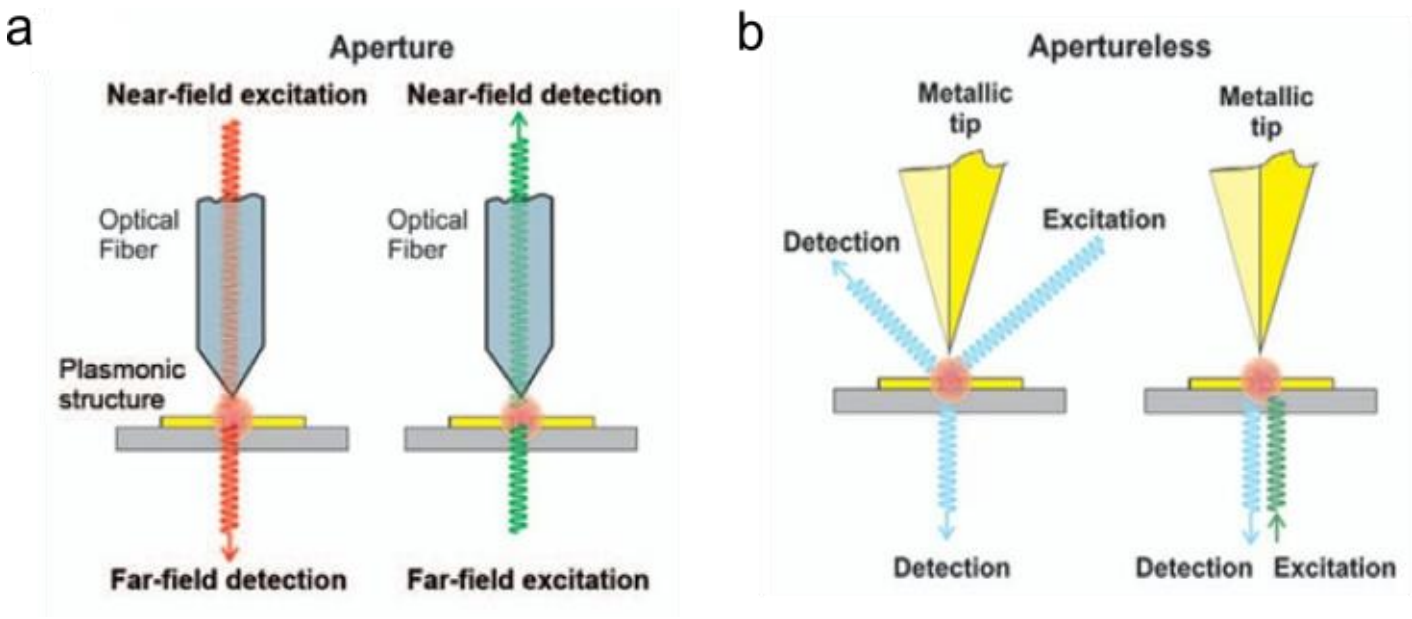

Fig. 31. Schematic of NSOM techniques using optical fiber probe ("with aperture") or metallic tip ("apertureless") [110].

For the study of plasmonic nanostructures, a variety of optical signals can be measured using NSOM, such as fluorescence, luminescence, second or third harmonic generation, etc. Fig. 32 shows an example of topography and experimental, background-free, near-field images of a rod, dish, and triangle antenna, which were excited close to their fundamental dipolar resonance at wavelength 9.6 μm [111]. [111]. The optical images were obtained using interferometric detection, dielectric tips, higher harmonic demodulation, and residual background subtraction. Both amplitude and phase images can be acquired using these configurations.

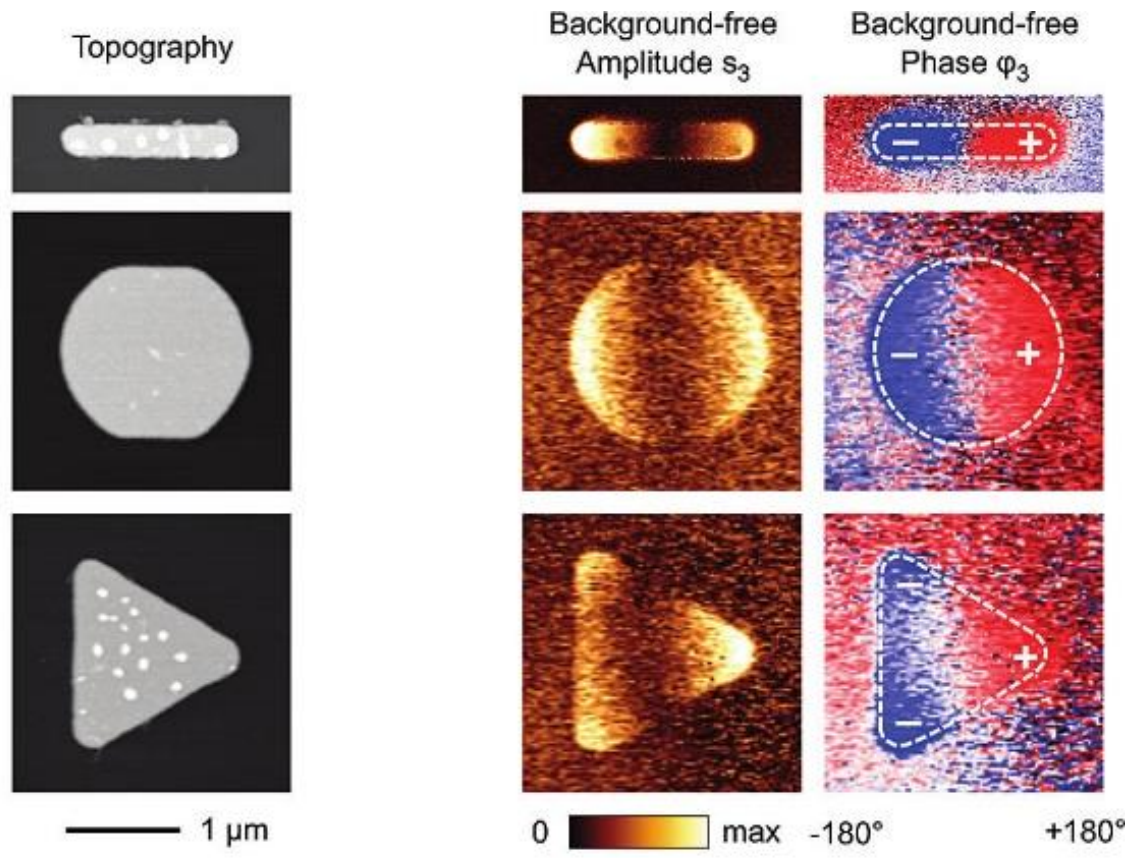

Fig. 32. Topography and experimental, background-free, near-field images of a rod, dish, and triangle antenna excited close to their fundamental dipolar resonance at $\lambda$=9.6 μm wavelength [111].

We have carried out a series of work in creation novel light source and nanoprobes for NSOM [19, 112, 113]. The large-scale applications of conventional NSOM are limited due to the manual assembly process of the probe with fiber and the requirement of an external light source. We presented a novel scanning "nanophotonic" microscope through monolithic integration of a nanoscale light-emitting diode (nano-LED) on a silicon cantilever. The nano-LED methods do not require external light source, and can reduce the size of light source to improve the resolution. Two types of nano-LEDs have been successfully developed using nanofabrication and MEMS technologies: 1) the fabrication of thin silicon dioxide light-emitting layer between heavily doped p+ and n+ silicon layers, fabricated by a focused ion beam and 2) electrostatic trapping and excitation of CdSe/ZnS core-shell nanoparticles in a nanogap. These probes have been employed into a standard near-field scanning and excitation setup. The probe successfully measured both optical and topographic images of chromium patterns with imaging resolutions 400 and 50 nm, respectively (Fig. 33) [112].

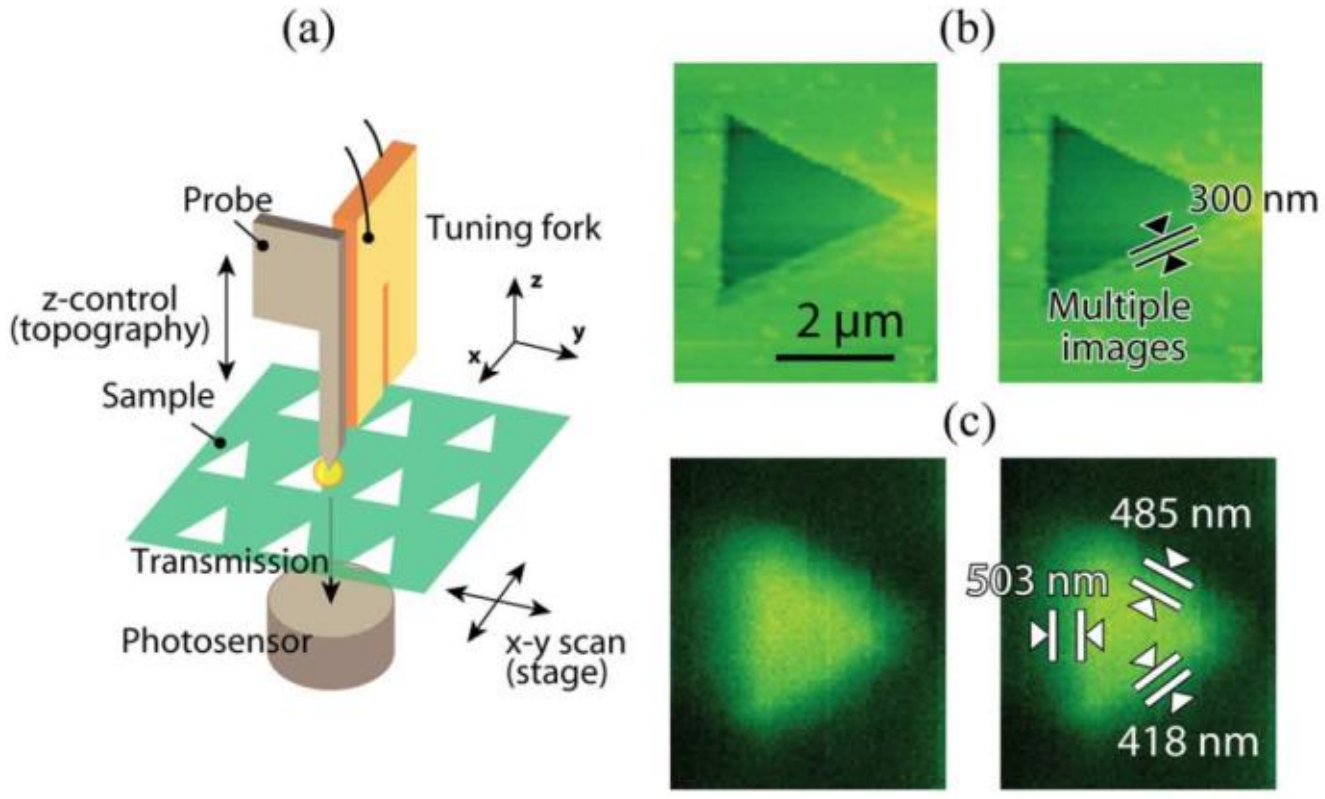

Fig. 33. NSOM using QD LED [112]. (a) Schematics of a standard experimental setup. (b) Topographic and (c) optical images of the chromium pattern on a glass substrate.

Compared to the topographic resolution of 50 nm, the relatively coarse optical resolution of 400 nm can be further improved through QD-plasmonic coupling and plasmonic enhancement [114]. With the capability from near UV to IR emission, plasmonic enhanced, mass-producible, and self-illuminating nano-LED on-chip will open up many exciting opportunities in biomedical and industrial applications, including near-field microscopy of subcellular structures, direct material patterning, and compact "light-on-chip" biosensors and biochips. Plasmonic structures including plasmonic microplates [115], thin metal films [116], gold colloids [117], plasmonic wells [114], and plasmonic nanogratings [19] have been applied to enhance luminescence of QDs.

NSOM probes have been used commonly in plasmonic enhancement for ultrahigh biosensitivity and single molecule detection [118, 119].

Two well-known techniques exploiting near-field plasmonic enhancement for ultrahigh sensitivity are "tip-enhanced infrared spectroscopy" and "tip-enhanced Raman spectroscopy".

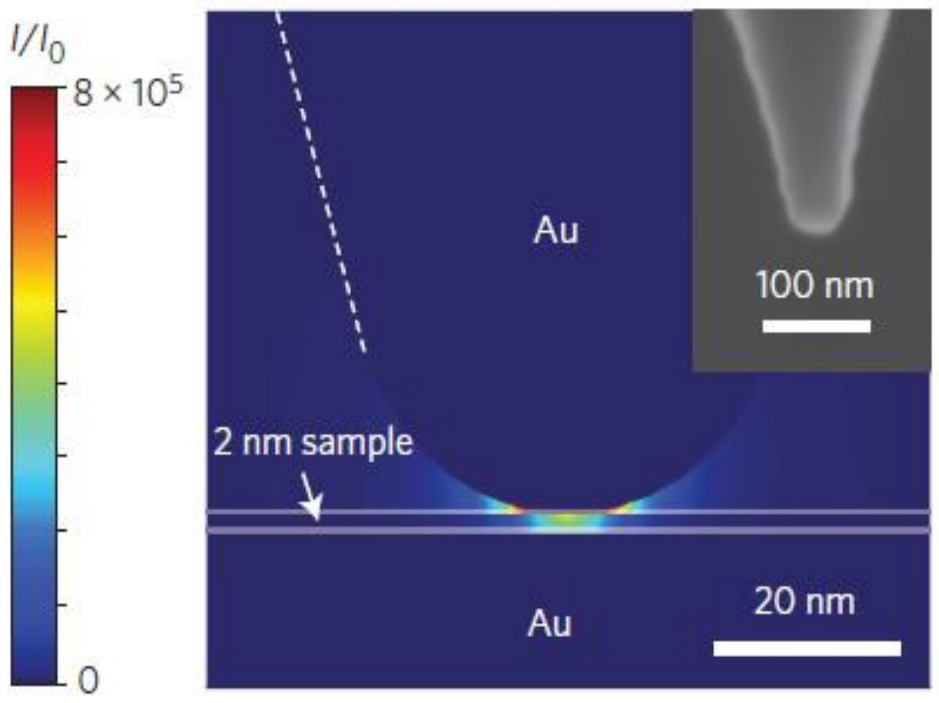

Fig. 34. Numerical simulation of the tip-enhancement of the light intensity (I/I0) for a 2-nm-thick SAMs on gold. The mid-infrared light is incident at 75 degrees to the surface normal. The AFM tip has the curvature radius of 25nm and half-cone angle of 17º, SEM of the actual geometry (inset).

Tip-enhanced infrared (TEIR) spectroscopy combines infrared spectroscopy with an AFM tip to enhance and measure the local absorption of IR light with specimen. Obtained absorption spectra can be used to investigate chemical and biological samples.

Belkin's group in 2014 utilized the TEIR technique to identify the nanoscale mid-infrared spectra of self-assembled monolayers (SAMs) by measuring the molecular expansion forces [120]. First the sample was illuminated via an IR light source which excites vibrational states of molecules. As a result, volume expansion of sample happens due to collective oscillation of vibrational modes (thermal expansion). This expansion leads to a slight cantilever deflection which can be detected using a position-sensitive photodetector. A gold coated AFM probe was exploited to enhance the vibration states via near-field plasmonic enhancement. Fig. 34 shows the utilized gold coated AFM tip used to enhance the IR absorption.

In a notable and exciting example of near-field plasmonic enhancement for nanospectroscopy and biosensing, authors in [121] have demonstrated that scattering-based NSOM (s-NSOM) is applicable to analyze and sense biological samples in aqueous media. Fig. 35 shows the set up used in their experiment. Two unique and exceptional properties of graphene, including the impermeability and IR plasmonic effect of graphene have been used to achieve live cell and virus sensing in aqueous solutions.

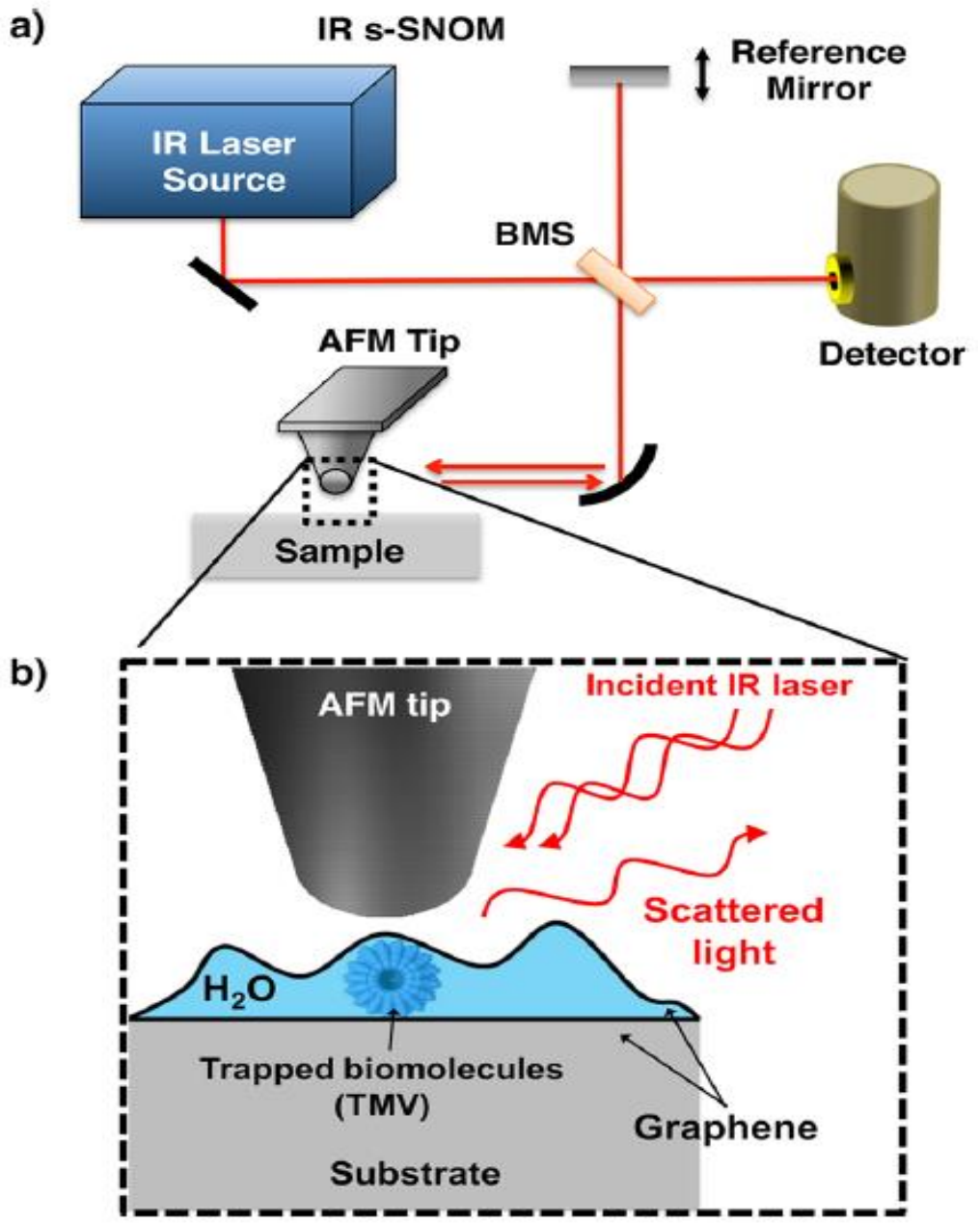

Fig. 35. Setup used in the wet s-NSOM. (a) Schematic of the setup, (b) tip-sample interaction [121].

Chemical vapor deposition (CVD) prepared monolayer graphene functions as an impermeable layer which reflects the light and allows sensing of the underlying molecular materials in an aqueous environment. The second graphene layer on mica substrate is used for plasmonic enhancement of vibrational stats in biomolecules at the interface.

In contrast to TEIR in which molecules absorb incident photons that are characteristic of their structure and match the vibrational frequency, in tip-enhanced Raman spectroscopy (TERS) molecules are firstly excited to a so-called virtual state and then relaxed back to the ground state, emitting photons with energies equal to the vibrational quantum [122]. Vibrational spectroscopy based on TERS can provide a valuable "fingerprint" of molecules by exploiting near-field surface plasmon enhancement at the tip apex. A comprehensive review on current advances in TERS can be found in [123, 124]. Here we briefly review recent works for biosensing applications.

Some of the earliest biomolecules to be studied by TERS were nucleic acids [125]. In one of the pioneering works, Deckert's group used TERS to provide a spectrum of the pyrimidine bases (thymine and cytosine) which then was used for DNA sequencing. In several reports, TERS has been utilized for amino acids sensing. As an example, authors in [126] presented the spectra of a short-immobilized peptide on a single transparent gold nanoplate where the gold plate provided a gap-mode plasmonic enhancement.

One the most attractive TERS results was recently reported by the Dong research group [127]. Using plasmon-enhanced Raman, the group was able to realize sub-nanometer (below one nanometer) spatial resolution of porphyrin, resolving its inner structure and surface configuration. The key to achieving such an ultrahigh-resolution was to enhance the vibrational modes of single molecules by pacing the them under an STM nanocavity plasmon and spectrally match the resonance of the nanocavity plasmon to the molecular vibronic transitions.

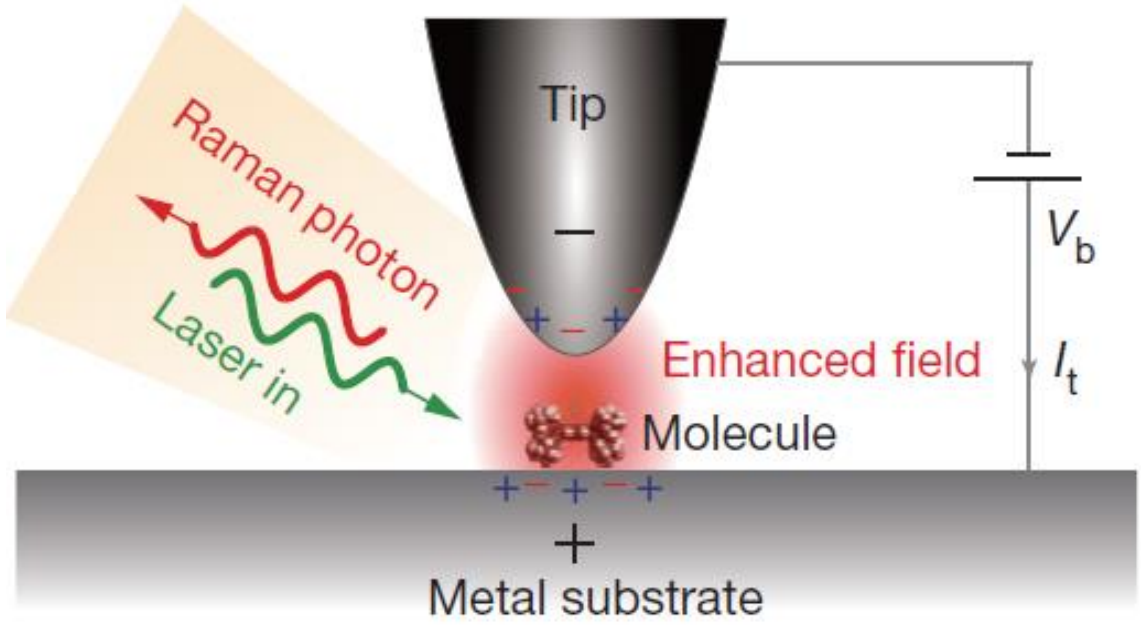

Fig. 36. Schematic of STM-TERS. $V_b$ is the sample bias and $I_t$ is the tunneling current.

Besides the fervent and intellectual demonstration of sub-nanometer single molecule sensitivity, this work opened up new fundamental questions about theoretical and physical principles of this phenomenon.

Raschke group utilized adiabatic plasmon concentration and grating-coupling to concentrate optical energy to the apex of an NSOM probe to suppress background signal in TERS [128]. The plasmonic grating was fabricated via FIB on the tip shaft as shown in Fig. 37. The extension of TERS into near-infrared was demonstrated due to adiabatic plasmon nanofocusing.

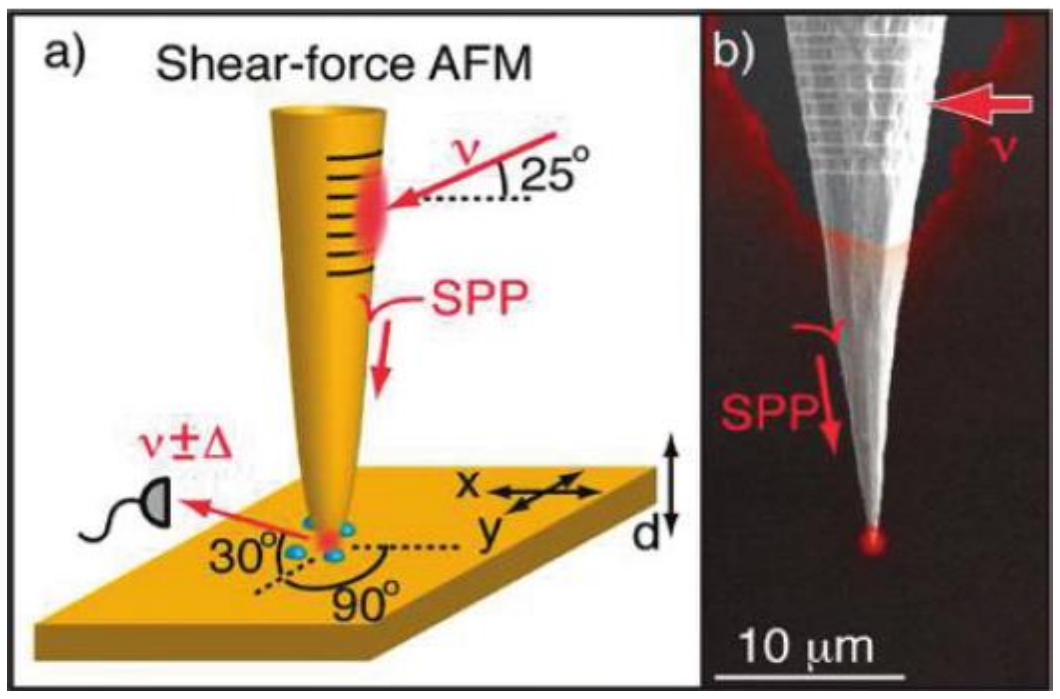

Fig. 37. (a) Schematic of the setup used in the experiment. (b) SEM of the probe. Gold tips are electrochemically etched and are mounted onto a quart fork. Plasmonic nanograting is realized via FIB [128].

### *D. Plasmonics-enhanced LED chip for cellular imaging*

Various microscopy techniques have been developed to provide sensitive imaging for cell observation. Fluorescence microscopy is one of the highly sensitive detection techniques. Fluorescence imaging of cells generally requires a bright image with high contrast and space resolution for immediate cell diagnosis. Therefore, plasmonic structures can be applied to enhance the cellular imaging, either as a substrate or as a light source [129, 130]. Plasmonic substrates have been used for high-resolution cellular imaging as well as monitoring the interaction of cells with their extracellular matrix. In one example, a plasmonic dish composed of a grating substrate was integrally molded with a cultivation plastic dish, and was applied to enhance the fluorescence imaging of cells [129].

The results showed fluorescence images of human embryonic kidney (HEK) cells were above 10 times brighter than those obtained on a conventional glass-bottomed dish (Fig. 38). In addition, to demonstrate the biocompability of the substrate, neuronal cells were successfully culture on the plasmonic dish for 10 days. The brighter fluorescence images with higher contrast could be obtained by controlling the thickness and flatness of metal layers. In another example, a nanocone substrate on silicon coated with silver was used for surface-plasmon enhanced fluorescence detection and 3D cell imaging [130]. The plasmonic substrate supported several plasmonic modes that can be coupled to a fluorophore, which can be used to enhance the fluorescence signal from both cell membrane and cytoplasm. 3D fluorescence enhancement was also observed. The fluorescence intensity from the fluorophores bound on cell membrane was amplified more than 100-fold as compared to that on glass substrate.

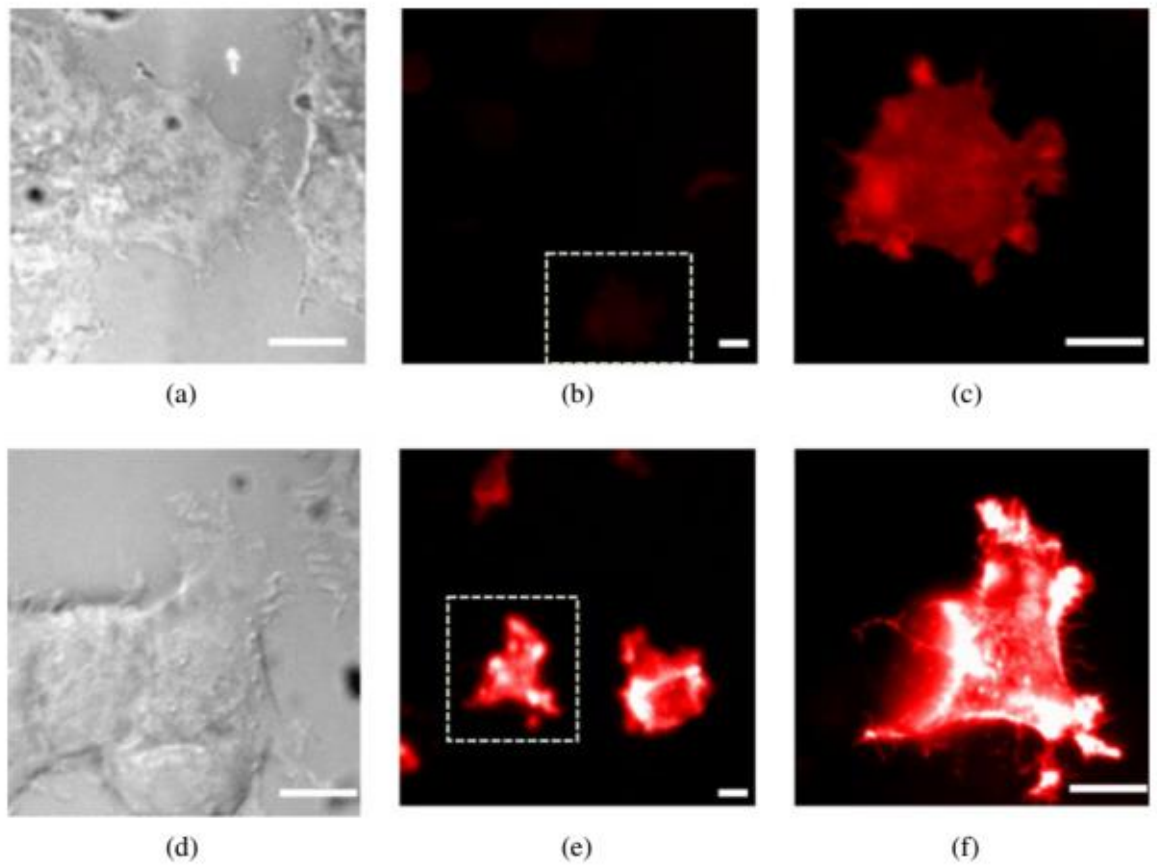

Fig. 38. (a, d) Bright-field images and (b, c, e, f) epifluorescence images of HEK cells observed on the glass slide (a-c) and plasmonic dishes (d-f). Scale bar 10 μm.

LEDs have been used as light source for cellular imaging in miniature systems. Since the wavelength depends on the energy bandgap structure, commercial LEDs are limited with the unavailability of narrow wavelength bands. Therefore, filters and other elements are required to obtain the demanded wavelength band for imaging, which increase the cost and size of the miniature system. Plasmonic quantum dot (QD) LEDs provides an excellent solution by combining the compact size of LEDs and the desirable wavelength of QDs. The unique characteristics of QDs, including tunable emission wavelength, narrow bandwidth and the capability of photo- and electrical-excitation, makes them ideal material as on-chip light source for cellular imaging. Bhave et al. integrated a 2D plasmonic nanogratings structure with QD LEDs as a compact light source for on-chip imaging of tumor cells [19].

Fig. 39a and 39b show the schematic of the light source structure and the set up used for cell imaging, respectively.

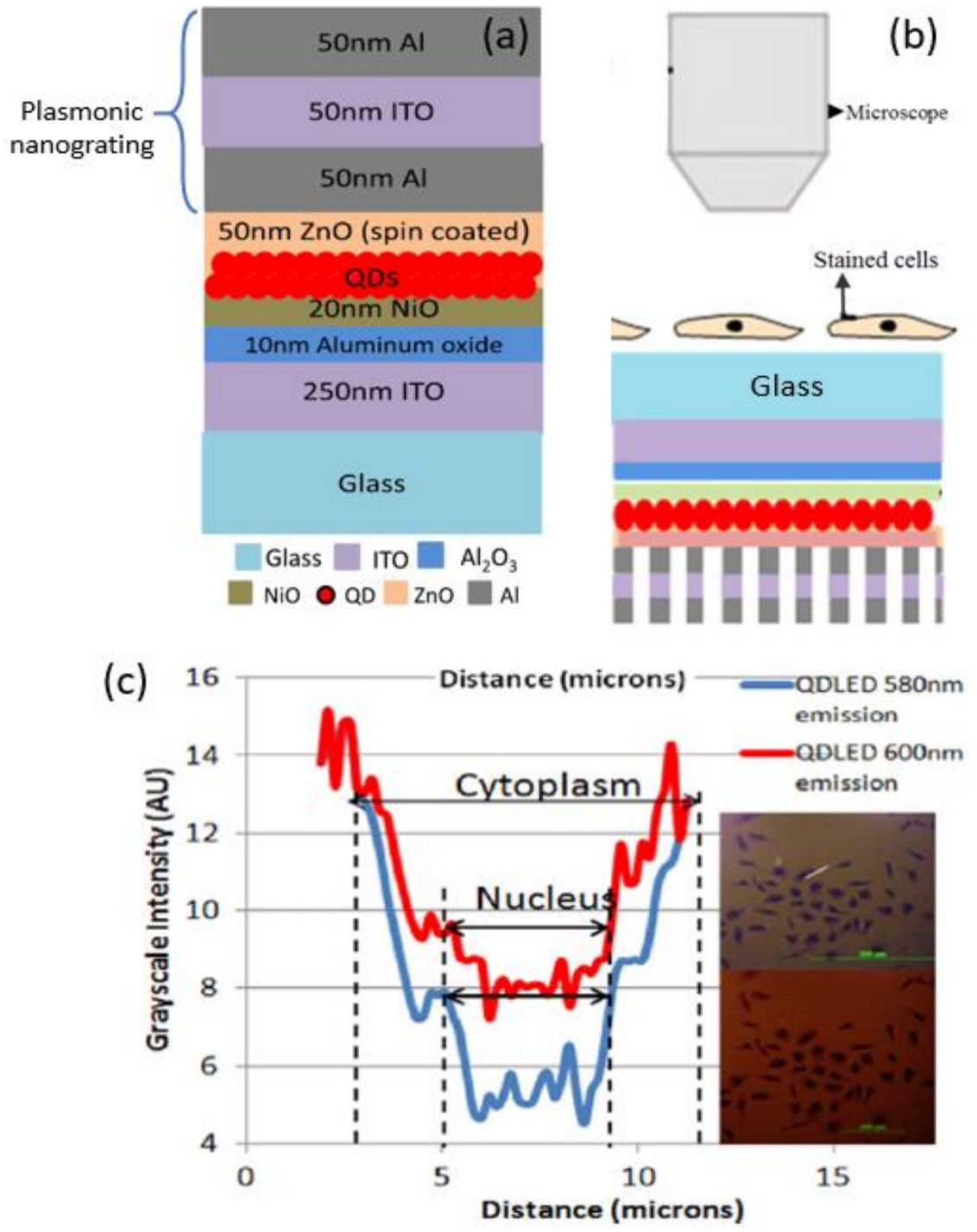

Fig. 39. (a) Schematic of the plasmonic nanogratings integrated with QD LED. (b) Schematic of setup used for cancer cell imaging using QD LEDs. (c) Cell imaging with electroluminescent QD LED sources. Inset shows the cell image with QD580 and QD600 sources; the plotting is the intensity profiles for grayscale-converted images from the pixels covered by the line in each of the images [19].

In this study, QDs were used as the emission layer, while the nanograting was employed to enhance the light intensity through the resonant reflection of surface plasmon waves. Fig. 35a shows the schematic of the light source structure. The parameters of nanogratings, including periodicity, slit width and thickness of the metal and dielectric layers, were designed to tailor the frequency bandgap to matches the operation wavelength. There was an increase of 34.72% in electroluminescence intensity, and 32.63% in photoluminescence. *Ex vivo* transmission microscopy was performed to evaluate the nucleus-cytoplasm ratios of MDA-MB 231 breast cancer cells using the QD LED light source, and the wavelength dependent imaging of different cell components was shown in Fig. 39c.

In addition to the examples presented above, more emerging concepts, designs and applications of plasmonic microsystems have been demonstrated, ranging from plasmonic gas and chemical sensors [131, 132], microfilters for circulating tumor cells (CTC) capturing [133], Surface-enhanced Raman spectroscopy (SERS) [134-137], single-molecule dielectrophoretic trapping [138-140], DNA biosensing [141] among many others. Fascinating new concepts are explored in detail in the framework of this review, with the goal to devise new geometries on microsystem surfaces that may for example require no moving structures to tilt the radiated beam in the desired direction. We predict that the combination of design theory, modeling and experimental implementation may provide full degrees of freedom and enhanced performance for future plasmonic microsystem designs for specific applications.

## V. Perspectives

There are synergies in advancement in plasmonic theory with microsystems design is emerging, where fascinating research has been carried out to perform imaging, sensing and light harvesting with unprecedented tempo-spatial resolution and throughput. We reviewed the selected applications concerned with plasmonic patterns on surfaces for biosensing applications. It illustrates the key points that this paradigm of plasmonic-on-chip enables low-profile conformal surfaces on microdevices, rather than a bulk material or coatings, which may provide clear advantages for the physical, chemical and biological-related sensing, imaging, light harvesting applications in addition to significantly easier realization, enhanced flexibility and tunability.

It is likely that the current trend in plasmonic microsystems design and implementation will contribute to multifunctional tasks to be integrated on a single chip [142, 143]. As presented in our previous work, tunable beam manipulation and biosensing can be realized on a single device to link near-field sensing with far-field detection [50]. The plasmonic structure presented in [121] consists of gold bowtie nanoantenna arrays fabricated on $SiO_2$ pillars which can be employed as a direct laser writing and plasmonic optical tweezers. It is worth noting that a combination of traditional MEMS with plasmonic sensing can be realized [144-149]

## VI. Conclusion

In this article, we identified the emerging trend of synergizing plasmonic patterns on microstructures towards advanced sensing, imaging, and energy harvesting. We reviewed the theory, fabrication, and application of plasmonic microsystems and the recent progresses in the field for biosensing applications. We believe that the optical plasmonic patterns on surface concept may constitute the much sought-after flexible and reliable bridge between near-field sensing, imaging at the nanoscale and far-field detection. In our vision, these concepts may be combined to realize a fascinating paradigm to manipulate light at will, with a clear impact on several key areas of interest for MEMS in addition to significantly easier realization, enhanced flexibility, and tunability.

## Acknowledgement

This work was partially supported by National Institute of Health (NIH Director’s Transformative Research Award, No. R01HL137157), National Science Foundation (NSF CAREER Award Grant No. 0846313) and DARPA Young Faculty Award (N66001-10-1-4049). We appreciate the helpful discussions with Professor Andrea Alu at the University of Texas at Austin, Professor Stephen D. Gedney at the University of Colorado Denver and Professor Kazunori Hoshino at University of Connecticut. The authors thank Amogha Tadimety for insightful editing of the paper.